\documentclass[11pt]{article}

\input epsf.sty

\usepackage{color} 
\usepackage[utf8]{inputenc} 

\usepackage{geometry}
\usepackage{graphicx}
\usepackage{xcolor}
\usepackage{amsmath}
\usepackage{amsfonts}
\usepackage{amssymb}
\usepackage{ascmac}
\usepackage[T1]{fontenc}
\usepackage{tikz}
\usetikzlibrary{positioning,calc}
\usepackage[compat=1.1.0]{tikz-feynhand}
\usepackage[colorlinks, citecolor=blue, linkcolor=black]{hyperref}

\usepackage{graphics,amsmath,amssymb}

\def\lromn#1{\uppercase\expandafter{\romannumeral#1}}

\definecolor{red}{cmyk}{0,1,0.50002,0}
\definecolor{blue}{rgb}{0,0.49995,1}
\definecolor{green}{cmyk}{1,0,0.49998,0}
 
\begin{document}
\begin{flushright}
\today \\
\end{flushright}

\begin{center}
\begin{large}
\textbf{
Constraints on extended Jordan-Brans-Dicke gravity
}

\end{large}
\end{center}

\vspace{1cm}

\begin{center}
\begin{large}

Kunio Kaneta\footnote{
kaneta@het.phys.sci.osaka-u.ac.jp,
 Present address: Department of Physics,
Osaka University},
Kin-ya Oda\footnote{odakin@lab.twcu.ac.jp}, 
and Motohiko Yoshimura$^{\dagger}$\footnote{yoshim@okayama-u.ac.jp} 

\vspace{0.5cm}
Department of Mathematics, 
Tokyo Woman's Christian University, \\
Tokyo 167-8585, Japan

\vspace{0.5cm}
$^{\dagger}$
Research Institute for Interdisciplinary Science,
Okayama University, \\
Tsushima-naka 3-1-1 Kita-ku Okayama
700-8530, Japan

\end{large}
\end{center}

\vspace{3cm}

\begin{center}
\begin{Large}
{\bf ABSTRACT}
\end{Large}
\end{center}

Cosmological analysis of 
 extended Jordan-Brans-Dicke (eJBD) gravity is presented 
in the Einstein metric frame in which gravitational interaction is readily understandable.
Our formulation is the first systematic investigation of how  to
introduce lagrangian of standard particle physics
in eJBD framework consistently with the general principle of spontaneously broken
gauge symmetry, which makes it possible to 
confront eJBD-based cosmology
with observational and laboratory bounds on time variation
of parameters, masses, and coupling constants,
caused by time evolution of eJBD fields.
Decomposition of standard particle physics lagrangian into 
independent gauge invariant pieces is
proposed to avoid  serious conflict that may arise from standard
lagrangian transformed from the Jordan frame.
Independent conformal factors  are assigned to each of
five gauge invariant pieces.
The formulation is most unambiguously made possible by 
defining fields having canonical kinetic terms
that allow us to use the canonical quantization rule of field theory.
This construction gives as one of its consequences
the canonical eJBD field $\chi$ that couples to
the universal fermion current, a linear combination of
baryon and lepton number currents, 
$\partial_{\mu} \chi (3  j_B^{\mu} + j_L^{\mu})$,
in addition to the conventional  trace of the energy-momentum tensor.
Field equation of eJBD field along with gravitational equation is analyzed by using a simplified
polynomial class of potential and conformal functions,
giving time evolution of radiation, matter and dark energy densities
consistent with observations when an appropriate set of model parameters
are used.
Finite temperature  corrections are further calculated to give
temperature dependent terms in eJBD field potential.

\vspace{2cm}

Key words

Dark energy,
Quintessence,
Jordan-Brans-Dicke gravity,
Finite temperature effects,
Time varying fine structure constant

\setcounter{footnote}{0}
\newpage
\section
{\bf Introduction}

Scalar field is an essential ingredient to inflationary universe models.
Moreover, it  may also be important for solving
the dark energy problem.
Among many scalar-tensor gravity theories,
 extended Jordan-Brans-Dicke~\cite {jbd} (abbreviated by eJBD in the present work) theories
belong to a promising class, as recently discussed in many disguises
of inflationary models \cite{ejbd} and of dark energy models
\cite{quintessence recent, dark energy cosmology: recent, symmetron model, 
many others}.

Historically, the scalar field,  in addition to the tensor field, was first introduced
in gravity by Jordan and Brans-Dicke (JBD) \cite{jbd}.
JBD theory was extended to a more general class of
scalar-tensor theories incorporating essentially two arbitrary functions of scalar field
by Bergmann and Wagoner \cite{bergmann, wagoner}.
We use in this work the terminology of extended JBD theory
(eJBD) for this class of extension of JBD scalar-tensor gravity.
Depending on the choice of two introduced functions,
one can describe a wide class of scalar-tensor gravity.
Popular quintessence model \cite{quintessence}
belongs to eJBD theory by choosing a single power or single exponential 
function and by introducing
a cosmological constant in the Jordan metric frame.

eJBD theory is classified into two types.
Type \lromn1 models have a potential in which the eJBD field settles down
to a finite field value at recent epochs, 
hence resulting in general relativistic cosmology
with a finite cosmological constant, namely
it effectively gives  $\Lambda$CDM model at the present epoch.
Chameleon and related models using a different
potential given in \cite{dark energy cosmology: recent, symmetron model}
belong to this class.
There remains a serious problem of fine-tuning cosmological constant
in some of these models.
On the other hand, type \lromn2 model makes the field never to settle,
approaching the field infinity asymptotically, if one takes
the asymptotic ratio of two introduced
functions  satisfies a condition.
The original quintessence model \cite{quintessence}  
belongs to this class of models as well.
A merit of our models is that it may open the possibility of solving the fine-tuning problem
of cosmological constant, as suggested for instance in \cite{cc relaxation}.

It is important in our view that the eJBD field varies with cosmic time evolution
in order to explain the smallness of dark energy of order (meV)$^4$
at present \cite{cc relaxation}.
We are thus led to concentrate on type \lromn2 eJBD models
in the present work.
Time variation of the eJBD
is called the  {\it field sliding}\footnote{ We prefer to use this terminology
instead of often used {\it runaway} to emphasize a slow time variation of field.} 
in the present work, when the field monotonically
and slowly increases towards infinity.
A similar idea of dark energy field is quintessence \cite{quintessence}
where a simple exponential or a power potential of dark energy field
is introduced.
We choose a different kind of potential, since our ultimate goal
is to attribute the same eJBD field to the origin of inflation as well.
The choice of field potential to realize the sliding
is limited, but the freedom allowed in the present new formulation
of eJBD gravity makes it natural to choose an interesting class of potentials.

There are a number of important issues to accommodate standard particle
physics lagrangian in eJBD gravity when one applies eJBD gravity
to cosmology.
We first raise   the question of metric frame difference in cosmology
based on eJBD gravity.
There have been confusing statements which one of
Jordan or Einstein frames is physically more relevant
in cosmology \cite{metric frames, metric frames 2, metric frames 3}\footnote{
It has been shown  in \cite{frame equivalence} that the Einstein
and Jordan frames are physically equivalent when all the higher mass dimensional
terms are included. The apparent discrepancy is due to the difference
of frames in which the action is assumed to take a particular, simplified form.
}.
We take a practical view that the Einstein frame is
more convenient in treating gravitational interaction.
Furthermore, a metric frame should be preferably
defined to give a constant time clock basic to clarify the space-time structure. 
The clock is ordinarily 
defined by using atomic energy difference to give a precise and stable
period of oscillation. Hence we find it
best to search for the Einstein frame that provides a constant electron inverse-mass,
and the constant fine structure constant if possible, for a typical clock.
The metric frame thus defined assumes the gravitational
constant $G_N$, the electron mass $m_e$, $\hbar$,  and the speed of
light $c$ to be time invariants.
We may however allow other constants such as the fine structure constant,
the proton mass, the W-boson mass, and the Higgs coupling
to vary with cosmological evolution, if such is allowed in an acceptable framework.

We formulate the question of parameter time variation 
such that independent gauge invariant pieces of standard particle physics lagrangian
are regulated by independent powers of a universal conformal function
of eJBD theory in the Jordan frame.
Quantum field theory of standard particle physics
is then subjected to various cosmological constraints
on time variation of masses and coupling constants typically.
Recently derived observational (and laboratory) constraints are very severe, and we find
eJBD theories that pass all of these tests to be limited.
For instance, if one naively introduces standard particle physics
lagrangian in the Jordan metric frame and
introduces the Weyl transformation to change it to the Einstein frame,
the derived lagrangian is found inconsistent with, hence is excluded by,
stringent bounds on variation of the fine structure constant.

Stringent bounds on time varying parameters exist.
These observational bounds are derived from cosmology limited to epochs
at and after big-bang nucleo-synthesis (BBN).
Thus for the analysis of this problem, the behavior of eJBD field 
as dark energy component becomes important only at these later cosmological epochs.
It is however possible that the eJBD field provides important clues for outstanding problems
of cosmology at earlier epochs such as the generation of baryon asymmetry and
inflation.
Discussion of cosmology prior to BBN requires how the  eJBD field
evolves, including finite temperature effects.
We initiate in the present work to work out finite temperature effects
in eJBD theory,
and to clarify the nature of electroweak spontaneous symmetry breaking 
using our new formalism.

Cosmological  mass variation in eJBD theories has been noted by many people
in the literature \cite{variable mass}, who proposed to use this variation
for their own different purposes.
Theoretical motivations in favor of time varying coupling constants, or
dimensionless mass ratios, have been reviewed in \cite{uzan}.
But to the best of our knowledge
there exists no systematic investigation of how one can
introduce mass and coupling time variations in eJBD-based gravity
consistently with  general quantum field theoretical principles of standard particle
physics.
We hope that our general approach in the present work provides a basic framework
of discussing time variation of parameters in eJBD gravity.

Work closest to ours in its spirit is \cite{damour-polyakov}.
String theory connection to dilaton-matter coupling
 is emphasized in this work, while
our work takes a more pragmatic view concerning the principle
of introducing particle physics into eJBD gravity,  restricting the gauge invariance
and renormalizability of standard particle physics.
Our canonical formalism is more general than the way how
the standard particle physics is introduced 
under the name of {\it least coupling principle} in \cite{damour-polyakov}.
Moreover, this work insists on the massless JBD field, which is
finally anchored at a stationary point in the recent cosmological epoch,
while our work considers a more general  massive eJBD field
still time varying at present.

The new contributions of the present work are to work out allowed couplings
of extended Jordan-Brans-Dicke scalars to standard model.
This is done consistently with known wisdom of super-string theory 
\cite{damour-polyakov} and
spontaneously broken gauge symmetry of standard model.
Identification of this class of general scalar-tensor theories
is important to determine the parameter range 
that is constrained by the non-observation of time variation of
fine structure constant.
In the literature \cite{broken scale inv} a subset of these theories is used to discuss
the fifth-force type of constraints, but there are other cosmological
constraints not discussed in that work as well.
One needs to fully incorporate standard model lagrangian
for a complete discussion of observational constraints, including all five independent gauge invariant pieces  as done in the present work.

The present paper is organized as follows.
In Section 2 we explain why we encounter serious problems
in eJBD gravity that contradict with observational bounds
on time variation (null result) of standard particle physics parameters.
We next introduce the idea of how one can modify eJBD theory
without any conflict to the basic concept of spontaneously broken gauge theory
of particle physics. Appendix A explains how one proceeds to
derive a canonical form of such extension, making readily usable
the  canonical quantization rule  of  field theory.
In this section we summarize results derived in  Appendix A.
It is pointed out that the canonical eJBD field has a  new
derivative coupling to the universal fermion current of
$ (3 j_B + j_L)^{\mu}$.

In Section 3 eJBD scalar dynamics is explored.
This is necessary to discuss theoretical predictions of
time variation against observational bounds, and the
canonical form of eJBD theory helps greatly.
Some analytic solutions are presented, which make easier derivation
of allowed model parameter regions from confrontation with observational bounds.
Numerical computations are also presented to show how appropriate parameter
choice leads to time evolution of radiation-, matter-, and dark- energy densities,
and of the equation of state factor after BBN consistent with
observation and theoretical prediction.
The technique explained in Appendix B is used for this numerical simulation.
In Section 4 we discuss how five observational and laboratory bounds
on time variation of the fine structure constant, a mass ratio and
null result on violation of equivalence principle constrain
model parameters in our canonical approach.
All these bounds have been obtained from observations
and theoretical calculations at BBN epoch and 
more recent epochs.

Section 5 starts discussion on earlier epochs of cosmology,
and examine how electroweak symmetry breaking may be modified.
We further discuss in Section 6 finite temperature effects, and 
calculate finite temperature corrections to the kinetic energy of eJBD field
as well as potential energy.
Finally, in two  Appendices we present systematically how we arrive at
the canonical form of eJBD gravity coupled to standard particle physics
in Appendix A and how we numerically compute time evolution
of each component of energy densities, radiation, matter and dark energy,
along with the equation of state factor in Appendix B.

Parameter constraints have been widely discussed in many different
kinds of scalar-tensor gravity.
For an extensive review of constraints on coupling constant variation
and theoretical motivations in favor of the variation, we refer to \cite{uzan}.
Our discussion of cosmological constraints on model parameters
is based on the new general formulation of eJBD theories not restricted
to specific models.
Brief discussion on the constraints are given in some references 
of \cite{dark energy cosmology: recent, quintessence recent}.
We shall cite other references relevant to our constraint later in 
Section 4.

Material presented in the present work contains
a versatile set of new possibilities, whose 
detailed ramifications are left to our future works of early cosmology.
In the present paper we provide the basic formalism
and its consistency with already existing observations.
Only a sketch of possible new avenues is described.

We use the natural unit of
$c=1 = \hbar$ unit  and the Boltzmann constant $k_B =1$,
unless otherwise stated.

\vspace{0.5cm}
\section
{\bf 
Standard particle physics coupled to canonical  eJBD field
}

We first point out a serious problem that one encounters if
one naively introduces lagrangian of standard particle physics into
eJBD gravity.

\subsection
{\bf Problem of eJBD gravity}

We take, as a theory of gravity, extended class of Jordan-Brans-Dicke scalar-tensor
gravity rather than general relativity.
The scalar-tensor sector of lagrangian in these theories is given by
\begin{equation}
\sqrt{-g} {\cal L}= \sqrt{-g}\,\left(
- \frac{1}{16\pi G_N} F(\phi) R + \frac{1}{2} 
 (\partial \phi)^2
-  V(\phi)  + {\cal L}_{st}\right) \,.
\label {original ejbd}
\end{equation}
For simplicity we took a single component eJBD field $\phi$.
Extension to multi-component eJBD fields is mentioned later.
${\cal L}_{st}$ is the textbook lagrangian density of standard particle physics
which does not contain the eJBD field $\phi$ at all.
Difference from general relativity is in the existence of a conformal function $F(\phi)$
multiplied to the Rici scalar curvature $R$,
and a potential function $V(\phi)$ which is absent in the original JBD theory \cite{jbd}.

The Einstein metric frame is obtained by
the metric scaling, called Weyl scaling, $g_{\mu\nu} \rightarrow F(\phi) g_{\mu\nu}$.
The resulting Einstein frame lagrangian is
\begin{eqnarray}
&&
\hspace*{-0.5cm}
\sqrt{-\bar{g}}  {\cal L}^{\rm E}(\bar{g}_{\mu\nu},\phi) = \sqrt{-\bar{g}} 
\left(- \frac{1}{16\pi G_N} R(\bar{g}_{\mu\nu})
+ 
\left( \frac{3 }{16\pi G_N } (\partial_{\phi}  \ln F)^2 
+ \frac{1}{2 F} \right) (\partial \phi)^2 
- \frac{1 }{F^2} V(\phi) 
 + \frac{1}{F^2}  {\cal L}_{st} (\psi, \frac{\bar{g}_{\mu\nu}}{F} )
\right)
\,.
\nonumber \\ &&
\end{eqnarray}
The power of $F$ factor in this equation
 is determined by the presence of metric $\bar{g}^{\mu\nu}$;
$F \bar{g}^{\mu\nu}$ for each $ \bar{g}^{\mu\nu}$
(or equivalently $\bar{g}_{\mu\nu}/F$ for each $ \bar{g}_{\mu\nu}$).
This naive manipulation leads to time variation of the fine structure constant obeying $\alpha/F^2$
\footnote{
As will be discussed later, the final form of time varying fine structure constant 
is more complicated than shown here, but its time variation is nonetheless strongly
constrained.
}, 
which is restricted
 by severe observational constraints, unless
eJBD field or $F$ stays nearly constant at late times
(this result is made evident when we later explain observational bounds
in modified eJBD theories).\!\footnote{
Field redefinition in conformal invariant theories eliminates time variation
of the fine structure constant
at the tree level, but conformal anomaly and quantum corrections at higher orders
of perturbation theory makes the time variant constant inevitable.
This time variation may be incorporated by an appropriate choice of $F$ factor
(including the possibility of non-universal F functions as discussed in the rest of the paper).
}

We thus need extension of eJBD framework to accommodate
standard particle physics\!\footnote{
At the tree level of string loop expansion one has a JBD lagrangian
$\sim e^{- 2\phi} ( -R + 4 (\partial \phi)^2) + {\cal L}_{st}$ \cite{string-motivated dilaton},
which is however drastically modified by string loop effects \cite{damour-polyakov}.
}.
Our proposal is to reformulate the rule of how to introduce standard particle physics:
we decompose ${\cal L}_{st}$ into five gauge invariant pieces and assign each of them
to separate independent $F$ factors \cite{stronger gravity}.
This way we respect the basic idea of spontaneously broken gauge invariance, 
allowing  more freedom to eJBD theory.
In future it may or may not become possible to justify
this independent assignment of conformal factors,
starting from superstring theory in higher spacetime dimensions
and compactifying it to four dimensional spacetime.
We shall  take a more pragmatic and phenomenological approach in the present work.

\subsection
{\bf How to introduce standard model in eJBD gravity}

It is practically important to define quantum field theory in terms of
canonical fields with which canonical equal-time commutation and anti-commutation
rules can be set up for bosons and fermions.
This can be done by a series of field re-definition.
Other parts of lagrangian, interaction and spontaneously generated mass terms,
are left intact, and we allow coupling constants and masses to vary with time.
This way one can use the technical machinery of quantum field theory 
along with easier comparison with observation bounds.
We derive in Appendix A canonical fields that have the standard
kinetic terms, $(\partial \phi)^2/2$
for bosons and $ \bar{\psi} i  \gamma \cdot \partial \psi$ for fermions.

Here is a summary of canonical formulation derived in Appendix A.
Standard particle physics is constructed from
four independent gauge invariant pieces of lagrangian, the gauge field ${\cal L}_g$,
the fermion field ${\cal L}_f$, the Higgs field ${\cal L}_H$,
the Yukawa coupling ${\cal L}_Y$. In addition there is a potential for eJBD field.
It is natural to expect \cite{damour-polyakov}
that string theory introduces five independent conformal
functions to these lagrangian operators; $\sum_i F_i(\chi) {\cal L}_i$.
We maximally utilize the freedom of field redefinition to transform these
into a standard form of canonical quantization.
When we write practically formulas in the Einstein frame, we specialize to 
the Friedmann-Lemaitre-Robertson-Walker (FLRW) metric given by
$ds^2 = dt^2 - a^2(t) (d\vec{x})^2$ \cite{cosmology textbook}.

\begin{itemize}
    \item Gravity and eJBD scalar sector: the action given by
   \begin{align}
&&
        S
=
        \int d^4x\sqrt{-g}\left[
        -\frac{M_{\rm P}^2}{2}R 
+ \frac{3}{4}M_{\rm P}^2g^{\mu\nu}\partial_\mu\ln F_{\rm g}\partial_\nu\ln F_{\rm g}
+\frac{1}{2}\frac{F_{\rm{d}\phi}}{F_{\rm g}}g^{\mu\nu}\partial_\mu\phi\partial_\nu\phi-\frac{V_{\rm J}(\phi)}{F_{\rm g}^2}
    \right],
   \end{align}
with $M_{\rm P} = 1/\sqrt{8\pi G_N} \sim 2.4 \times 10^{18}$GeV the
reduced Planck mass.
The term $ - \frac{3}{2}M_{\rm P}^2g^{\mu\nu}\nabla_\mu\nabla_\nu \ln F_{\rm g}$
that exists in the original form is a surface term, and deleted in this formula.
    The relation between $\chi$ and $\phi$ depends on $F_{{\rm d}\phi}$.
    The canonically normalized modulus field $\chi$ should be defined as
    \begin{align}
        \frac{1}{2}g^{\mu\nu}\partial_\mu\chi\partial_\nu \chi
        &\equiv
        \frac{3}{4}M_{\rm P}^2g^{\mu\nu}\partial_\mu\ln F_{\rm g}\partial_\nu\ln F_{\rm g}+\frac{1}{2}\frac{F_{\rm{d}\phi}}{F_{\rm g}}g^{\mu\nu}\partial_\mu\phi\partial_\nu\phi.
    \end{align}

The rest of standard model Einstein frame lagrangian is given in terms
of rescaled canonical fields.
    \item Gauge sector:
    \begin{align}
        {\cal L}_g &=
        -\frac{1}{4}F^{\mu\nu}F_{\mu\nu} - \sum_{a=1}^3\frac{1}{4}W^{a\mu\nu}W^a_{\mu\nu} - \sum_{A=1}^8\frac{1}{4}G^{A\mu\nu}G^A_{\mu\nu}.
    \end{align}
    \item Fermion sector
\footnote{  
With $\partial_{\mu} \ln F_{\rm df}= F_{\rm df}^{-1} \partial_{\chi}F_{\rm df} \, \partial_{\mu} \chi$ 
in the following formula
(\ref{derivative fermion coupling}), the eJBD field has a derivative coupling to
fermions $\propto \partial_{\mu} \chi$.
The existence of this important fermion coupling is overlooked
in the work of \cite{damour-polyakov}.
Its derivation is given in Appendix 8.4 and originates from
the formula.
}:
    \begin{align}
        {\cal L}_f &=
        \sum_\psi
        \overline\psi i\gamma^\mu \left(
        \nabla_\mu + \frac{1}{2}\partial_\mu \ln F_{\rm df}
    \right)\psi,
\label {derivative fermion coupling}
    \end{align}
    where
    \begin{align}
        \nabla_\mu &=
        \partial_\mu + \frac{1}{4}\omega_{\mu ab}\gamma^{ab} - ig_{Y,{\rm eff}} Y_\psi A_\mu - ig_{2,{\rm eff}}W_\mu - ig_{3,{\rm eff}} G_\mu 
    \end{align}
    with the effective gauge coupling $g_{A,{\rm eff}}=F_{\rm dA}^{-1/2}g_A$.
    Note that $(1/4)\omega_{\mu ab}\gamma^{ab} = \frac{3}{2}\dot{a}\gamma^0/a$.
    \item Higgs sector:
    \begin{align}
        {\cal L}_H &=
        g^{\mu\nu}(D_\mu H)^\dagger (D_\nu H) - 
 V_{\rm eff}( H) ,
    \end{align}
    where
    \begin{align}
        V_{\rm eff}( H)  &= F_{\rm g}^{-2}V(H)- g^{\mu\nu}\left\{
        (\partial_\mu\ln f_{\rm dH})(\partial_\nu\ln f_{\rm dH})+2\nabla_\mu\nabla_\nu\ln f_{\rm dH}
    \right\}|H|^2
\label {higgs sector canonical}
    \end{align}
    with $f_{\rm dH}^2\equiv F_{\rm dH}/F_{\rm g}$.
The potential $V(H)$ is of the usual textbook wine-bottle shape that
is used in general relativistic cosmology.
The second contribution of this equation has an interesting structure,
which may be recast to
\begin{eqnarray}
&&
- \left( 3 (\partial_{\chi} \ln f_{dH})^2 \partial^{\mu}\chi \partial_{\mu} \chi
+ (\partial_{\chi} \ln f_{dH} ) \nabla^{\mu}\nabla_{\mu} \chi \right) |H|^2
\,.
\label {non-trivial higgs pot-correction 2}
\end{eqnarray}
This way we derive the identical result to the one given in \cite{broken scale inv}
when taking $ F_{dH}  \rightarrow 1$ limit.
Under the FLRW
 metric the coefficient of the second $|H|^2$ term reduces to
\begin{eqnarray}
&&
- 3 (\partial_{\chi} \ln f_{dH})^2 \dot{\chi}^2 - \partial_{\chi} \ln f_{dH} (\dot{\chi}^2
- 3 \frac{\dot{a}}{a} \dot{\chi})
\,.
\end{eqnarray}
    \item Yukawa coupling part:
    \begin{align}
        {\cal L}_y &=
        -y_{u,{\rm eff}} \overline Q H^c u - y_{d,{\rm eff}} \overline Q H d - y_{e,{\rm eff}} \overline L H e - y_{\nu,{\rm eff}} \overline L H^c \nu_R + \cdots ,
    \end{align}
    where the effective Yukawa coupling is given by ($f= u, d, e, \cdots$)
    \begin{align}
        y_{f,{\rm eff}} &= F_{\rm Y}F_{\rm df} F_{\rm dH}^{-1/2} y_f.
    \end{align}
\end{itemize}
We assumed in this formula Dirac-type neutrino masses,
but Majorana-type neutrino masses can be accommodated by
introducing SU(3)$\times$SU(2)$\times$U(1) singlet $N_R = \nu_R$
with their Majorana mass terms.

Besides these, remnant effects of Jordan frame lagrangian appears in
the gauge and Yukawa couplings:
\begin{eqnarray}
&&
g_{A,{\rm eff}}=F_{\rm dA}^{-1/2}g_A
\,, \hspace{0.3cm}
 y_{f,{\rm eff}} = F_{\rm Y}F_{\rm df} F_{\rm dH}^{-1/2} y_f
\,.
\end{eqnarray}
The effective Yukawa coupling $y_{f,{\rm eff}} $ here shall be
reflected in fermion mass variation via time varying $\chi$.

\vspace{0.5cm}
Different conformal functions $F_i$ that appear in our general formalism give too much of 
unnecessary complications, and there is no cosmological data
to sufficiently constrain these.
We shall propose 
as a pragmatic compromise to use different powers of a common conformal 
function $F$ described by a simple polynomial of $\chi^2$.
In particular, we parametrize as follows:
\begin{eqnarray}
&&
    F_{\rm g} = F^{\eta/2}\,, \hspace{0.3cm} 
    F_{\rm df} = F^{p_{\rm df} } \,, \hspace{0.3cm} 
    F_{\rm dA}^{-1/2} = F^{p_g}\,, \hspace{0.3cm} 
 \frac{F_H}{F_{\rm g}^2} = F^{p_H} \,,
\\ &&
    f_{\rm dH}  = \sqrt{\frac{F_{\rm dH}}{F_{\rm g}}} = F^{p_{\rm dH}} \,, \hspace{0.3cm} 
    F_{\rm Y}F_{\rm df} F_{\rm dH}^{-1/2} = F^{p_Y}\,,
\end{eqnarray}
so that we may write
\begin{eqnarray}
&&
V_{\rm eff}(\chi)  = \frac{V_{\rm J}(\chi)}{F^\eta} \,, \hspace{0.3cm} 
\frac{i}{2}(\partial_\chi \ln F_{\rm df}) (\partial_\mu\chi)\overline\psi\gamma^\mu \psi 
= i\frac{p_{\rm df}}{2}\frac{\partial_\chi F}{F} (\partial_\mu\chi)\overline\psi\gamma^\mu \psi
\,,
\\ &&
 g_{A,{\rm eff}} = F^{p_g}g_A \,, \hspace{0.3cm} 
 y_{f,{\rm eff}} = F^{p_Y}y_f
\,.
\end{eqnarray}
Our proposal of eJBD field coupling to standard model particles
gives a non-universal coupling, different from (\ref{original ejbd})  in the Jordan frame.
As discussed in the next section, this is necessary to evade 
cosmological constraints given there.

Relation of canonical eJBD field and original eJBD field in the Jordan frame
can be worked out explicitly for a single component eJBD scalar,
as discussed in Appendix A.

In summary, we write canonically normalized massive eJBD field $\chi$ coupling 
to five independent gauge-invariant  pieces in the Einstein metric frame\!\footnote
{The assumption of
{\it least coupling principle} in \cite{damour-polyakov}
excludes the possibility of some non-vanishing
powers $p_i$ in this formula.
We allow all these powers by taking the view that they are only bounded by
observations. In Section 4 it is found that only $p_g$ and $p_Y$ are restricted
by observations.
}:
\begin{eqnarray}
&&
{\cal L}^{\rm E}_{st} =  F^{p_g}{\cal L}_g + F^{p_{\rm df} } {\cal L}_f -
F^{p_H} V_{\rm J}(H) +   F^{p_{\rm dH}} (D_{\mu} H)^{\dagger} (D^{\mu} H)
+ F^{p_Y}{\cal L}_y
\,.
\label {standard model in e-frame}
\end{eqnarray}

There are two new functions, $V_J(\chi)$ and $F(\chi)$, in our simplified
extended Jordan-Brans-Dicke gravity.
We shall take the following simplest functions for these, as fully explained
in the next section:
\begin{eqnarray}
&&
F = 1 +  f (\frac{\chi}{M_{\rm P}})^2 \,, \hspace{0.3cm}
f > 0
\,, \hspace{0.5cm}
V_{\rm J} = V_0 \left( d_4 (\frac{\chi}{M_{\rm P}})^4 
 + d_2 (\frac{\chi}{M_{\rm P}})^2  + 1  \right)
\,, \hspace{0.3cm}
 d_4 V_0 > 0
\,.
\label {conformal and potential functions}
\end{eqnarray}
Equations, (\ref{standard model in e-frame}) and (\ref{conformal and potential functions}),
define our simplified eJBD theory, with additional term of (\ref{non-trivial scalar-coupling}).

Notice that in this simplified framework, the physical constants depend on time through the six parameters, namely, $\eta, p_{\rm df},p_g,p_H,p_{\rm dH},p_Y$, two of which will be 
severely constrained in the present work.

Non-trivial terms that cannot be given by conventional terms
times $F-$power functions emerge in the Higgs sector (\ref{higgs sector canonical})
and  the fermion sector having a new structure, 
\begin{eqnarray}
&&
\frac{i p_{\rm df} }{2}(\partial_{\chi} \ln F)\, \partial_\mu \chi 
\sum_{\psi}   \overline{\psi }\gamma^\mu \psi 
\,, \hspace{0.5cm}
\sum_{\psi}   \overline{\psi }\gamma^\mu \psi = 
 3j_B^{\mu} + j_L^{\mu}
\,,
\label {non-trivial scalar-coupling}
\end{eqnarray}
with $j_B^{\mu}, j_L^{\mu}$ the baryon number and the lepton number current, respectively.
In addition to this derivative coupling, eJBD field couples with
the trace of energy-momentum tensor,
 $\propto \sum_f m_f \bar{f} f$ after the electroweak symmetry breaking,
as usual for the dilatonic coupling.

The current that appears in the derivative coupling (\ref {non-trivial scalar-coupling})
is universal to all fundamental fermions, given by a linear combination
of baryon number and lepton number currents.
It was pointed out in \cite{cohen-kaplan} that
this form of derivative coupling can create a chemical potential
to distinguish matter from anti-matter, ultimately
leading to generation of a baryon number asymmetry
along with the field sliding,
when some new interaction violating either the baryon
or the lepton number is introduced as in grand unified theories.
Thus, it may have new important implications to lepto-genesis,
if the standard particle physics is extended to include heavy
Majorana particles, as suggested in \cite{lepto-genesis by sliding inflaton}.

The universal fermion current that appears in (\ref{non-trivial scalar-coupling}),
or its variant current, has been postulated in \cite{fayet} to couple to
a very light gauge boson.
There is an important difference of this scheme with ours:
our eJBD field is a scalar, instead of a vector,
 with mass varying as time evolves in cosmology. 
This difference makes it difficult to detect our eJBD field in
accelerator experiments.
Nonetheless, its extremely light mass at the present universe
leaves some common features with the gauge boson of \cite{fayet}
 such as test of the equivalence principle, which shall be discussed below.

\vspace{0.5cm}

\vspace{0.5cm}
\section {\bf Time evolution of eJBD scalar field}

We restrict our discussion to
 the spontaneously broken gauge theory of standard particle physics
in curved spacetime.
Quantum gravity is not necessary to consider unless one discusses the Planck epoch
in which the energy scale is of the order of $10^{18}$GeV.
For many cases of cosmological consequences one needs to consider
a limited class of the spatially flat  FLRW  metric.

\vspace{0.5cm} 
\subsection{\bf  Scalar field dynamics after BBN}

We shall work out the case of a single eJBD field\!\footnote{
The case of multi-component {\it N} scalar fields may be treated as follows.
We consider the multi-component eJBD gravity simplified
by imposing a symmetry to potential and conformal functions.
Take real {\it N} component eJBD theory as an example.
Simplification of scalar field equation is possible using O(N) symmetry
of {\it N}-component scalar fields for conformal function $F_{\rm g}$ and
potential $V$ \cite{my21-22}.
Let us take for illustration the simplest O(2) model 
that allows spontaneous symmetry breaking
(SSB) due to a wine-bottle type of potential.
Kinetic terms of two-component scalars may be written
modulus $\chi$ and angular $\theta$ modes:
\begin{eqnarray}
&&
\frac{1}{2}(\partial{\chi})^2 + \frac{1}{2}\chi^2 (\partial \theta)^2
\,.
\nonumber
\end{eqnarray}
Due to O(2) symmetry the angular mode does not appear in the potential term,
and it can be integrated by introducing two dimensional angular momentum $L$
as $\dot{\theta} = L^2 \chi^2/a^3 $.
This introduces an extra centrifugal term to the lagrangian for modulus field;
$L^2/(2 a^6 \chi^2)$.
Eigenvalues of non-Abelian O(N) group may be introduced for {\it N}-component scalar system
similar to this O(2) case.
},
to derive the lagrangian density:
\begin{eqnarray}
&&
 a^3 {\cal L}_{\chi} = a^3 \left( \frac{1}{2} \dot{\chi}^2 
- V_{\rm eff}(\chi) + \sum_{i} F^{p_i}(\chi) {\cal L}_i \right)
\,, \hspace{0.3cm}
V_{\rm eff}(\chi) = \frac{V_{\rm J}(\chi) }{F^{\eta}(\chi)}
\,,
\end{eqnarray}
where $  {\cal L}_i $ are five independent, gauge invariant pieces 
of standard particle physics lagrangian density.
Variational principle for  $\chi$ is worked out, to give differential equation, 
\begin{eqnarray}
&&
\ddot{\chi} + 3\frac{\dot{a}}{a} \dot{\chi} +  \partial_{\chi} V_{\rm eff}(\chi) = 
{\rm matter \; coupling\ contribution}
\,.
\label {scalar field eq rd}
\end{eqnarray}
The matter coupling contribution in the right hand side,
given roughly, but not precisely, by $(\rho - 3 p) \sum_i \partial_{\chi} F^{p_i} $
for ideal fluid of energy density $\rho$ and pressure $p$,
is given in necessary places below.
In the radiation dominant epoch this contribution is usually small
due to the traceless condition of matter energy-momentum tensor
 in the Einstein metric frame.\!\footnote{
It is shown in \cite{dark energy cosmology: recent}
that this is not necessarily so, because contributions of non-relativistic
particles at temperature $\sim $ particle mass provide kick to the scalar field,
a mechanism first pointed out in \cite{kick by nr matter}.
This type of chameleon effect that necessarily exists
in any eJBD model is however a minor contribution
in generic type \lromn2 models including our model
characterized by the small slope $\partial_{\chi} F^p$ from the following reason.
After the matter-radiation equality time, the background photon is the only major contribution in the right hand side. However, since $\chi$ does not couple to photons, the matter coupling contribution during the matter-dominated epoch is negligible. Therefore, we may safely neglect the matter coupling contribution for the entire evolution of $\chi$. 
The major part of cosmological time evolution in our model
is thus described without the matter coupling contribution.
But in discussion of cosmological bounds and finite temperature
corrections these contributions become important,
as explained in Section 4.
}

We now discuss how to choose the conformal and the potential functions, $F\,, V_{\rm J}$,
that appear in the effective potential $V_{\rm eff}(\chi) = V_{\rm J}/F^{\eta}$.
The guiding principle we adopt in the present work is the asymptotic form
of $V_{\rm eff}(\chi)$ at field infinity $\chi \rightarrow \infty$.
This should be a monotonically decreasing function to realize the field sliding
(sometimes called {\it runaway} in the literature),
and the simplest choice would be a negative fractional power as in
\cite{quintessence}.
It is impractical to choose infinite power series for $V_{\rm J}$ and $F$,
hence we truncate, for a pragmatic purpose, these functions to some finite positive powers, 
$p_j$ and $p_v$, respectively.
The asymptotic limit of the effective potential can become either increasing
(case of $p_v \geq \eta p_J +1 $) or decreasing (case of $p_v \leq  \eta p_J -1$), 
neglecting constant terms irrelevant to the potential force.
The case $p_v = \eta p_J$ has either of these behaviors, depending on more details of parameters.

We shall work out the simplest case 
of $p_v = 2, p_J=1\,, \eta=2 + \delta$,
with a small positive $\delta$ that give interesting results.
Larger maximum-power cases would work equally well,
however giving slightly more complicated time evolution at
intermediate epochs than the simplest case given below.
The potential $ V_{\rm eff}(\chi)$ has a small slope $\propto  \chi^{- 2\delta}$
at the field infinity for a small positive $\delta$.
We further restrict potential and  conformal functions
of renormalizable forms, to obtain
\begin{eqnarray}
&&
F = 1 + f_1 \frac{\chi}{M_{\rm P}}+  f_2 (\frac{\chi}{M_{\rm P}})^2 \,, \hspace{0.5cm}
f_2 > 0
\,,
\\ &&
V_{\rm J} = V_0 \left( d_4 (\frac{\chi}{M_{\rm P}})^4 + d_3 (\frac{\chi}{M_{\rm P}})^3
 + d_2 (\frac{\chi}{M_{\rm P}})^2  + d_1 (\frac{\chi}{M_{\rm P}}) + d_0  \right)
\,, \hspace{0.5cm}
 d_4 V_0 > 0
\,.
\label {f,v choice 0}
\end{eqnarray}
This form still contains too many parameters, and we further
impose discrete symmetry under $\chi \rightarrow - \chi$
for simplicity, to define standard potential and conformal functions,
\begin{eqnarray}
&&
F = 1 +  f (\frac{\chi}{M_{\rm P}})^2 \,, \hspace{0.5cm}
f > 0
\,,
\\ &&
V_{\rm J} = V_0 \left( d_4 (\frac{\chi}{M_{\rm P}})^4 
 + d_2 (\frac{\chi}{M_{\rm P}})^2  + 1  \right)
\,, \hspace{0.5cm}
 d_4 V_0 > 0
\,.
\label {f,v choice}
\end{eqnarray}
A potentially serious problem of overproducing domain walls
is ignored, since it is readily cured by introducing $\chi$ odd terms.

The large field limit of the standard effective potential is given by
\begin{eqnarray}
&&
V_{\rm eff} \approx d_4 f^{-2- \delta} V_0 y^{-2\delta} 
\,, \hspace{0.3cm}
y = \frac{\chi}{M_{\rm P}}
\,.
\label {eff potential at large field}
\end{eqnarray}
A more general choice of $F$ powers
satisfying $p_v = \eta p_J + 1$ gives similar, but not
identical, results
on cosmology after big-bang nucleo-synthesis.
Even more generally, models that give the asymptotic behavior
$ V_J/F^{\eta} \rightarrow O(\chi^{-2 \delta})\,, \delta > 0$
lead to similar results to ours.
Whenever this large field asymptotic behavior is maintained, 
one can derive similar cosmological constraints, which shall be later discussed.

The behavior near the field origin $\chi=0$ ($y=0$ too) is more complicated:
one derives by a power series expansion,
\begin{eqnarray}
&&
V_{\rm eff}(y) = V_0 \left( 1+ (d_2 - 2 f - \delta f) y^2
+\left\{ d_4 - d_2 f (2+\delta) + 3 f^2 + \frac{1}{2} \delta (5+ \delta) f^2
\right\} y^4 + O(y^6)
\right)
\,.
\end{eqnarray}
This potential form coincides with quintessence potential of power-law type
\cite{quintessence}, but our potential at smaller fields deviate from
the quintessence potentials.
The interesting case is provided by choosing a negative $y^2$ coefficient
and a positive $y^4$ coefficient, just as in the wine-bottle potential
in electroweak symmetry breaking.

\begin{figure*}[htbp]
 \begin{center}
 \epsfxsize=0.6\textwidth
 \centerline{\epsfbox{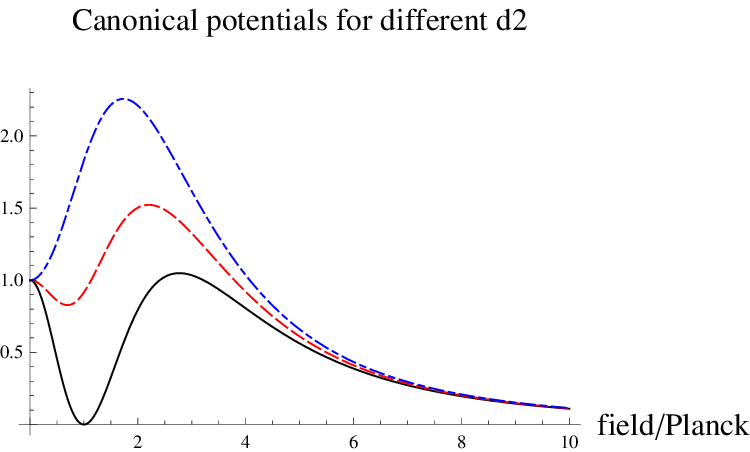}} \hspace*{\fill}\vspace*{1cm}
   \caption{
Potential shapes for a common set of $d_4=1, \eta =3.5 (\delta =1.5), f =0.25, V_0=1$
and $d_2=-2$ in solid black, 0 in dashed orange, and 2 in dash-dotted blue.
}
   \label {potential forms}
 \end{center} 
\end{figure*}

We illustrate typical potential shapes in Fig(\ref{potential forms}).
Introduction of $F$ powers in the denominator for our canonical
form of potential $V_{\rm J}/F^{2+\delta}$ drastically changes
the original potential shape given by $V_{\rm J}$, in the small field region.
This may give a variety of different physical consequences in cosmological applications
from the simple $V_{\rm J}$ case.
Nonetheless, the asymptotic behavior at the field infinity
has the same shape, which gives a unique late cosmology.
A typical case we adopt is a negative $d_2$ case as in the solid black curve.
Local extrema seen  in Fig(\ref{potential forms}) are not always
present, however, and may disappear to give
monotonically decreasing function of eJBD field, if one arranges
coefficients of both quadratic and quartic $\chi$ powers to vanish.
For a large coefficient $f$ the power series expansion given above is not useful,
and the asymptotic region may set in even when  fields are not very large.
To late-time cosmology relevant to the dark energy problem
 these $f$ dependences are not very important,
but they are important to inflationary epoch:
without local extrema it is difficult to realize inflation.
These matters shall be discussed separately.

A comment on a simpler choice of potential form may be
appropriate.
The simplest choice would be the exponentially decreasing
(or monotonically decreasing single-power)
function $V_Q$ often used in quintessence dark energy models
\cite{Copeland:1997et}.
Such models with a single exponential potential however suffer from producing overabundant dark energy, except for the case with carefully tuned parameters \cite{Franca:2002iju}.
This and $V_Q/F^2$ ($1/F^2$ also
monotonically decreasing) both give monotonic decrease towards larger fields,
and do not provide an interesting possibility to connect the eJBD field
to inflation.
For this reason we shall disfavor quintessence type of theories.
Our choice of potential and conformal functions gives
a variety of effective potential as illustrated in Fig(\ref{potential forms}).
Nevertheless, we may point out that some new findings in the present work
are valid for quintessence model as well.

\subsection
{\bf Analytic solutions}

A nice feature of this asymptotic potential is that
there exists a power-law solution in radiation dominated epoch of
$a \propto t^{1/2}$.
This solution is given by
\begin{eqnarray}
&&
\chi = M_{\rm P} D (M_{\rm P} t)^\beta
\,, \hspace{0.3cm}
\beta = \frac{1}{1+ \delta}
\,, \hspace{0.3cm}
D = \left( \frac{4\delta (\delta+1)^2 }{\delta + 3 } 
f^{-2 - \delta} \frac{d_4 V_0 }{M_{\rm P}^4 } \right)^{1/(2+ 2 \delta) }
\,.
\end{eqnarray}
The solution in matter-dominated epoch is similar:
the only change is in the coefficient $D$,
\begin{eqnarray}
&&
D \rightarrow D (\frac{2( \delta + 2)}{ \delta + 3} )^{1/(2+ 2 \delta) }
\,.
\end{eqnarray}

We illustrate a few examples of the effective potential thus defined
in Fig(\ref{potential forms}).
It is interesting that different magnitudes and signs of quadratic coefficient $d_2$ of the potential
$V_{\rm J}$ give qualitatively different behaviors.
This is understandable since this sign determines whether
the original scalar field potential admits a spontaneous symmetry breaking.
It was pointed out in \cite{cc relaxation}
that the change of symmetry breaking pattern, as seen from orange to blue curves
in Fig(\ref{potential forms}),  due to quantum correction at
finite temperature is a clue to solve the cosmological constant problem.

The Einstein equation for the cosmic scale factor is 
\begin{eqnarray}
&&
(\frac{\dot{a}}{a})^2 = \frac{8\pi}{3} G_N 
\left( F^{\epsilon_r} \, N\, \rho_r + F^{\epsilon_m} M n_M
+ \frac{1}{2} \dot{\chi}^2 + F^{-2 - \delta} V_{\rm J}
\right)
\,, \hspace{0.5cm}
\rho_r = a_B T^4
\,,
\label {einstein eq rd}
\end{eqnarray}
with $N$ the effective particle degrees of freedom regarded as
massless, $a_B$ radiation energy constant, and $n_M$ the number density of cold dark matter
of mass $M$.
We do not need to mention relation of $\epsilon_r\,, \epsilon_m$
to the powers we already introduced.
We assumed dark matter contribution behaving in the same manner as massive
standard model fermions.
The scalar field contribution is in the form of dark energy with
a constant cosmological constant $\Lambda$ replaced by
\begin{eqnarray}
&&
\rho_{\rm DE} =
\frac{1}{2}  \dot{\chi}^2 + F^{-2 - \delta} V_{\rm J}
\,, \hspace{0.3cm}
F^{-2 - \delta} V_{\rm J}
\approx d_4 f^{-2- \delta} V_0 (\frac{\chi}{ M_{\rm P}})^{-2\delta}
\,.
\end{eqnarray}

Analytic solution in the potential-dominated dark energy region
may be derived by dropping the potential force term in the scalar equation.
Solution has the form,
\begin{eqnarray}
&&
\chi = M_{\rm P} D_2 \, t^{1/\delta}\;
\,, \hspace{0.3cm}
D_2 = \left( \sqrt{\frac{3 d_4 f^{-2-\delta}  V_0 }{ (\delta -1 )M_{\rm P} }} \right)^{1/\delta}
\,, \hspace{0.3cm}
a(t) = (\frac{t}{t_0})^{(\delta - 1)/3}
\,,
\end{eqnarray}
under the condition $\delta > 1$.

$\chi-$field equation above includes its time variant mass and interaction
with matter fields.
The $\chi$ mass is
is estimated from (\ref{eff potential at large field}):
it is,
in the most of radiation-dominated epoch,
\begin{eqnarray}
&&
m_{\chi} \sim \sqrt{\frac{d_4 V_0} {M_{\rm P}^2} } 
\sqrt{2 \delta (2\delta + 1) f^{-2-\delta} y^{-2 - 2 \delta} } 
\,, \hspace{0.3cm}
\sqrt{\frac{ V_0} {M_{\rm P}^2} } = O(10^{-33}\, {\rm eV})
\,,
\end{eqnarray}
in the large field limit of background solution $y=\chi(t)/M_{\rm P} \propto t^{1/(1+\delta)}$,
hence very crudely $O(10^{-33}\, {\rm eV})\, (t_0/t)$.
The magnitude of this quantity is much smaller
than the electron mass even at the electroweak epoch,
and one may take $\chi$ field essentially massless.

Analytic solutions given above have a limited value due to
its small field range of applicability.
Hence we next present detailed numerical analysis.
Numerical analysis after BBN is discussed in detail in Appendix B in which our
methods of solving the set of differential equations
are also explained.
Results for radiation, matter, dark energy densities,
and the equation of state factor  are illustrated in Fig(\ref{time-evolution_fig1}) in the text
and Fig(\ref{time-evolution_fig2}) of Appendix B,
which show consistency with recent observations.
The equation of state factor and eJBD field in most recent era of $z= 100 \sim 0$
are illustrated for the same set of parameters in Appendix B. 
Note that eJBD field changes little:
its increase during recent epochs of $z+1 = 3 \sim 1$ is $\sim 0.025$,
as seen in Fig(\ref{time-evolution_fig2}).

\begin{figure*}[htbp]
 \begin{center}
 \epsfxsize=0.6\textwidth
 \centerline{\epsfbox{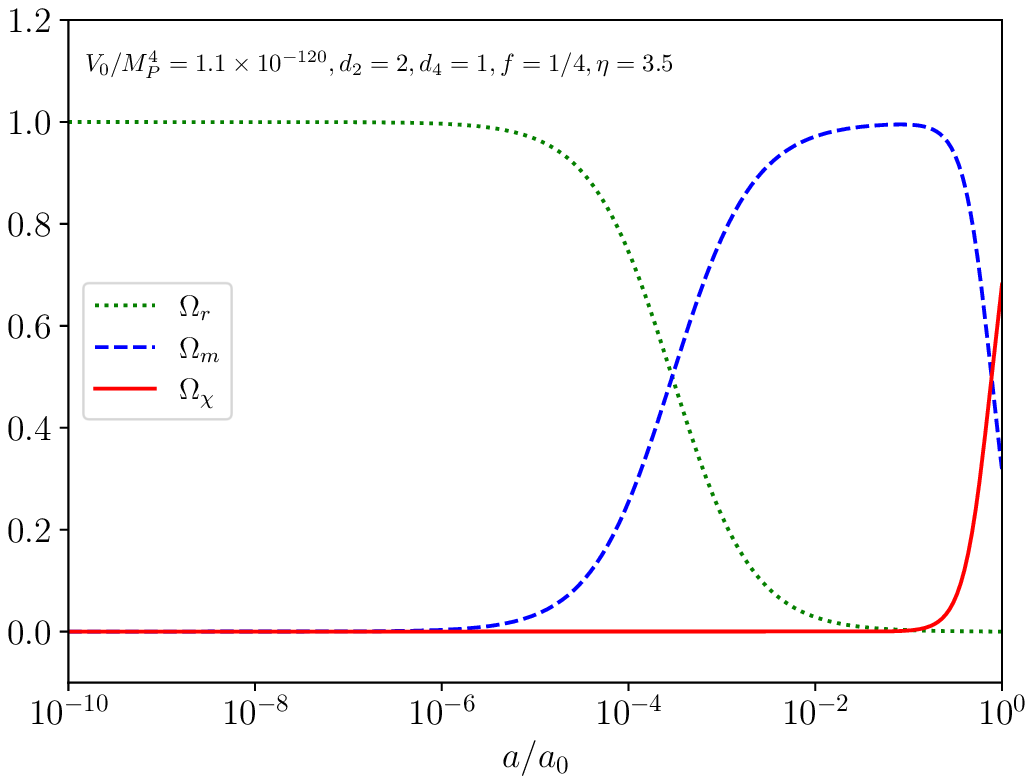}} \hspace*{\fill}\vspace*{\fill} \vspace*{-4cm}
   \caption{
Time evolutions of energy densities of radiation, matter, and eJBD field $\chi$
at epochs of $z= 10^{10} \sim 0$.
Assumed model parameters are $V_0/M_{\rm P}^4 = 1.1\times 10^{-120}$
(corresponding to $V_0 = (2.5 {\rm meV})^4$),  $ d_2=2, d_4 =1,
f=1/4, \eta=3.5(\delta = 1.5)  $.
}
   \label {time-evolution_fig1}
 \end{center} 
\end{figure*}

\vspace{0.5cm}
\section {\bf Confrontation with
 time variation bounds: cosmological observations
and laboratory experiment}

\subsection
{\bf Observational bounds from early universe}

We limit our discussions  to bounds on time variation of
fine structure constant, proton to W-boson mass ratio.
Our prediction of the effective fine structure constant is
\begin{eqnarray}
&&
\alpha_{\rm eff} = \alpha (\frac{t}{t_0})^{8 p_g/\delta}
\,, \hspace{0.3cm}
\frac{\delta \alpha_{\rm eff}}{\alpha_{\rm eff}} = 
\frac{8 p_g }{\delta } \frac{\delta t}{t}
\,,
\end{eqnarray}
where $p_g$ is the power of conformal factor in
the electroweak gauge sector, electromagnetic coupling given by $e F^{p_g}$.
The power for proton to W-boson mass ratio is more complicated,
as discussed below.

\vspace{0.5cm}
{\bf (A) Oklo bound}

Authors of \cite{oklo} improved the result of an earlier work
on the cosmological bound of $\alpha$ time variation,
and they conclude a bound,  $|\frac{\delta \alpha}{\alpha} | < O(10^{-7}) $.
Arguments for this constraint are based on the natural Oklo reactor
that happened $\sim 1.8$ billion years ago,
which is claimed to be sensitive to neutron capture cross section
for Sm$^{149} (n, \gamma)$Sm$^{150}$ proceeding via
resonance formation.

Actually, arguments for the bound given in \cite{oklo} are rather intricate.
A small resonance energy of order 100 meV arises from
a nearly complete cancellation of Coulomb and strong interactions
of protons inside Sm nuclei.
To make derivation of the bound meaningful,
the authors assume constancy of mass parameters like the proton mass.
They further assume that the strong coupling
constant does not vary with time, although 
some others \cite{alpha variation 2}  anticipate its variation.
Nonetheless, we take this as a reasonable bound on 
time varying strength of electrostatic interaction.

The theoretical bound on time variation from Oklo observation  is estimated as
$| p_g/\delta| < O( 10^{-7}) $
 using $\delta t /t_0 = 1.8/13.7 $.

\vspace{0.5cm}
 {\bf (B) Bound from quasar absorption lines}

Reference \cite{fs-splitting variation} searches for
time variation of $\alpha$ in fine structure splitting 
of alkaline-like atoms proportional to
a combination of parameters $(Z \alpha)^4 m_e $
\cite{atomic physics textbook}.
Observation data the authors use are from quasar absorption spectra.
Their result, $\delta \alpha/\alpha = (- 1.9 \pm 0.5) \times 10^{-5}$
for $z>1$,  may be compared with the predicted electrostatic
interaction of time variation.
The considered redshift factor range $\delta z/z \sim 0.6$ 
corresponds to $\delta t /t \sim 0.5$.
This gives A-scheme bound,
$| p_g/\delta| < 2.5 \times 10^{-6} $.
This constraint is  persuasive,
since the theoretical prediction is based on QED effects alone
without QCD effect (possibly except the minor effective mass factor 
$(1+ m_e/m_A)^{-1}$ with $m_A$ nucleus masses).

\vspace{0.5cm}
{\bf (C) BBN bound on $m_p/m_W$}

Present bounds on $G_F m_p^2 (\propto (m_p/m_W)^2)$
arises from BBN, \cite{BBN bound, gr tests}.
We first need to explain the origin of proton mass.

In order to estimate time variation of the proton mass $m_p$,
it is necessary to separate two parts: QCD gluon +
 quark kinetic energy and quark mass variations.
We take $m_p- 2 m_u - m_d$ for the first part of proton mass,
and the rest $2 m_u + m_d \sim 9.4\,$MeV for the second part.
In the canonical field formulation given in Appendix A,
gluon contribution does not have conformal factor,
while the quark mass part has a power dependence $F^{p_Y}$.
The ratio $r$ of the second to the first parts is $\sim 0.01 $ at the present epoch.
Gauge boson mass is controlled by another power $F^{p_g }$.
Dating back to earlier times, we model the mass ratio,
\begin{eqnarray}
&&
\frac{m_p}{m_W}(t) = \frac{m_p}{m_W}(t_0) F(t)
\,, \hspace{0.5cm}
F(t) = (1-r ) \frac{F^{ - p_g}(t) }{ F^{ - p_g}(t_0)}
+  r  \frac{ F^{p_Y - p_g}(t)}{ F^{p_Y- p_g}(t_0)}
\,.
\label {time variation mp/mw}
\end{eqnarray}
The predicted time variation of the mass ratio  is
\begin{eqnarray}
&&
\frac{\delta F}{F} = \left(
- \frac{4 p_g}{ \delta}(1 -r ) + \frac{ 4 (p_Y - p_g)}{\delta} \, r \right)
\frac{\delta t}{t}
\,.
\end{eqnarray}

The observational BBN bound 
 \cite{BBN bound, gr tests}
\begin{eqnarray}
&&
|\frac{d}{dt} G_F m_p^2 | < O(5 \times 10^{-12})\, {\rm yr}^{-1} 
\,,
\end{eqnarray}
is translated to parameter bound,
\begin{eqnarray}
&&
| - \frac{4 p_g}{ \delta}(1 -r) + \frac{ 4 (p_Y - p_g)}{\delta} \, r  |
< 0.7 \times 10^{-3}
\,.
\end{eqnarray}
If one combines the more stringent bound on $p_g/\delta$ from
the fine structure constant variation given above,
this gives effectively a bound, 
$|p_Y/\delta| < O(10^{-2})$.

\vspace{0.5cm}
\subsection
{\bf Scalar-exchange potential and equivalence principle}

The weak equivalence principle (WEP) that states equivalence of
inertial and gravitational masses has been tested with great precision.
For instance, the E$\ddot{{\rm o}}$tv$\ddot{{\rm o}}$s parameter
$\eta$ measured by MICROSCOPE mission \cite{microscope, gr tests}
is constrained by
\begin{eqnarray}
&&
|\eta( {\rm Ti, Pt})| < O(10^{-15})
\label {etovos}
\end{eqnarray}
The E$\ddot{{\rm o}}$tv$\ddot{{\rm o}}$s parameter is
difference of the inertial to gravitational mass ratio
between two isotopes.
We can attribute the attractive force caused by 
spatially inhomogeneous inflaton-modulus field $\delta \chi$
exchange to an addition to the inertial mass:
this is a kind of fifth force.

Let us discuss the law of gravity in the local Lorentz frame;
namely we assume a flat Minkowski metric at the present epoch.
Around the flat Mikowski metric of $\eta_{\mu \nu} =(1\,, -1\,, -1\,, -1)$ 
and the background scalar field configuration $\chi_0(t) $ of
$\chi = \chi_0\,, F_{\rm g}= F_0 = F_{\rm g}(\chi_0 )$,
one expands the metric in terms of second-rank tensor field 
$h_{\mu \nu}$ 
defined by $h_{\mu \nu} =g_{\mu \nu} - \eta_{\mu \nu}  $
and scalar field $\delta \chi = \chi - \chi_0$.
The resulting Einstein equation is that for massless graviton field $h_{\mu\nu}$
in which the source term contains both the usual matter-radiation
and the dark energy $\delta \chi$.

The scalar field equation is given by
\begin{eqnarray}
&&
 \partial^2  \delta \chi 
+ (\frac{\partial^2 V_{\rm eff}}{ \partial \chi^2})_0
\delta \chi=
\sum_{\psi} \left( \partial_{\chi} m_{\psi,{\rm eff}} \right)_{\chi = \chi_0} 
\langle \bar{\psi} \psi \rangle
\,,
\label {scalar eq}
\end{eqnarray}
to the leading Planck order.
This should be compared to general relativistic equation;
$\partial^2 h_{\mu\nu} + \cdots = -  \,T_{\mu\nu}^{(m)}/M_{\rm P} $.

In the large field limit of our eJBD model,
\begin{eqnarray}
&&
(\partial_{\chi}^2 V_{\rm eff} )_{\chi_0} 
\sim 2 \delta (2\delta +1) f^{-2 -\delta} \frac{d_4 V_0}{M_{\rm P}^2} 
(\frac{\chi_0}{M_{\rm P}} )^{-2 \delta -2}> 0
\,.
\end{eqnarray}
For a positive $\delta$ and $\sqrt{d_4 V_0}/M_{\rm P}$ of order
the present Hubble rate $\sim 10^{-33}$eV,
the $\chi$ quantum mass is exceedingly small,
and is taken as effectively massless.

The source term of scalar coupling to the trace of energy-momentum tensor is given by
\begin{eqnarray}
&&
\sum_{\psi} \left( \partial_{\chi} m_{\psi,{\rm eff}} \right)_{\chi = \chi_0} 
\langle \bar{\psi} \psi \rangle \approx f^{p_Y} (\frac{\chi_0}{M_{\rm P}} )^{2p_Y-1} 
\frac{M_A}{M_{\rm P}}
\,,
\end{eqnarray}
for a macroscopic massive body of mass $M_A$ (earth, astrophysical star etc).
If the power of conformal function 
$p_Y$ is equal to $ 1/2$,
the coupling strength of static scalar-exchange potential 
is of the same order as the gravitational strength.
If this power is less than 1/2, the strength may be much smaller
than the gravity.

Time variation of E$\ddot{{\rm o}}$tv$\ddot{{\rm o}}$s parameter is
\begin{eqnarray}
&&
\frac{2 (2 p_Y-1)}{\delta} \frac{\delta t}{t_0}
\,.
\end{eqnarray}
If this time span is 1 year, $\frac{\delta t}{t_0} \approx 0.7 \times 10^{-10}$,
and the pre-factor in this formula is restricted to be less than of order $10^{-5}$.
This bound is however applicable only when $p_Y$ is close to 1/2.
If $| p_Y| < 1/2$, one can safely ignore the experimental test of
equivalence principle.

\subsection
{\bf  Laboratory bounds derived from two compared atomic clocks} 

Ratio of hyperfine split (HFS) of $^{199}$Hg$^+$ transition to fine structure (FS)
split of alkali-like $^{27}$Al$^+$ transitions has been measured
over $\sim 1 $ years, and laboratory bound of the order, a few times
$ 10^{-17}$yr$^{-1}$ was derived \cite{atomic clock: alpha}.
FS is (special) relativistic effect of static Coulomb potential,
while HFS arises from magnetostatic interaction.
Since our formalism predicts constant electron mass and the
effective $\alpha $ variation, the formalism is applicable to this
transition ratio.
The bound is very useful, since other bounds, (A) and (B),  
necessarily involves assumptions on cosmology and geophysics,
which makes a quantitative evaluation of those bounds difficult.
The time span $\delta t/t_0 = 1/1.37 \times 10^{10}$ gives
a bound 
$| p_g/\delta| < 0.7 \times 10^{-7} $.

There are less crucial observation bounds that
are mentioned in a comprehensive review \cite{uzan}.

Summarizing all bounds discussed in this section,
the simplest way to clear observational and laboratory bounds is
the trivial power choice of $p_g =0\,, p_Y = 0$.
Even in this case, a non-trivial choice of $\delta > 0$
that appears in eJBD effective potential in the form $V_{\rm J}/F^{2+\delta}$  gives
interesting consequences to cosmology.
We mention two possibilities in the next two sections.

\section {\bf Electroweak symmetry breaking}

The spontaneous breaking of SU(2)$\times$U(1) gauge symmetry 
is one of the most successful parts of standard particle physics.
All masses of fundamental particles, leptons and quarks, W- and Z-gauge
bosons, and Higgs boson, are generated by this mechanism.
One might be concerned about how this success is changed
in eJBD gravity.

Discussion on the spontaneous symmetry breaking
starts from the  canonical form given by 
(\ref {non-trivial higgs pot-correction 2}).
Under the FLRW metric the canonical Higgs potential in eJBD theory is
\begin{eqnarray}
&&
\hspace*{-0.3cm}
 V_{\rm J}(H) = F^{p_H} 
V(H) + \Delta V(H)
\,, \hspace{0.3cm}
\Delta V(H) =
- \left(
3 (\partial_{\chi} \ln F^{p_{dH}} )^2 \dot{\chi}^2 + (\partial_{\chi} \ln F^{p_{dH}} ) (\ddot{\chi}
- 3 \frac{\dot{a}}{a} \dot{\chi}) \right) |H|^2
\,.
\label {higgs potential can-3}
\end{eqnarray}
Let us first concentrate on the second contribution.
We extrapolate the asymptotic field behavior
$\chi \propto t^{\beta}, \beta > 0$ back to electroweak energy scale;
with $\dot{\chi} = \beta \chi/t $ valid in the asymptotic region, one derives
the contribution of the second term in radiation-dominated (RD) epoch,
\begin{eqnarray}
&&
\Delta V(H) \approx 
- 2 \beta  p_{dH}  \,
\frac{ 6 \beta p_{dH} -\frac{\beta}{2} - 1 }{ t^2} 
 |H|^2
\,, \hspace{0.3cm}
\beta = \frac{1}{\delta + 1}
\,.
\end{eqnarray}

It is found that a crude and simple estimate of $\Delta V(H)$ is sufficient to
determine its importance.
Extrapolating RD even to the present gives the electroweak time
scale related to its redshift factor and the present Hubble value by a formula,
\begin{eqnarray}
&&
\frac{1}{t_{EW}^2} \approx H_0^2 z_{EW}^4 = O({\rm meV})^2
\,.
\end{eqnarray}
Correction of meV scale order to the Higgs potential is
entirely negligible, 
 and we shall ignore this contribution hereafter.

There is another related matter discussed in the next section, which
must be considered in a more systematic discussion of the electroweak phase transition.

Summarizing this section,
it seems necessary to reduce the power factor in front of Higgs potential:
using a common conformal function,
\begin{eqnarray}
&&
F(\chi) = 1 + f (\frac{\chi}{M_{\rm P}})^2
\,,
\end{eqnarray}
the Higgs potential is to a good approximation,
\begin{eqnarray}
&&
F^{p_H} V(H)
\,,
\end{eqnarray}
with a small value of $p_H \ll 1$.
Constrained powers of this common conformal function, 
$p_g, p_Y$, and unknown power $\eta = 2 + \delta$,
appear in front of gauge invariant pieces of lagrangian density.
Another power $p_f$ is introduced in the next section.

\section {\bf Dynamical temperature effects to eJBD 
field evolution
}

So far we limited our discussion to scalar field potential at zero temperature.
This approximation is justified when we concentrate on late time cosmology.
When we, however, discuss earlier cosmological epochs towards electroweak
scale of order 100 GeV, finite temperature effects may become important.
We shall discuss here 
finite temperature corrections in perturbation theory,
 mostly after the electroweak phase transition.

\subsection
{\bf Field coupling to fermions}

The eJBD $\chi$ field couples to standard model fermion field $\psi$ in the form,
\begin{eqnarray}
&&
{\cal L}_{\chi_0 \psi} =
 \frac{i p_f}{2} \partial_{\chi} \ln F\, \partial_{\mu} \chi\,
\bar{\psi} \gamma^{\mu} \psi
+  (F^{p_Y}(\chi) - 1 )\,    m_{\psi} \bar{\psi}  \psi
\,, \hspace{0.5cm}
\sum_{\psi} \bar{\psi} \gamma^{\mu} \psi = 
3  j_B^{\mu} + j_L^{\mu}
\,.
\label {fermion-chi coupling}
\end{eqnarray}
Field independent fermion mass terms are subtracted in this formula.
We have also suppressed other contributions  from Higgs boson and gauge boson.
The formula (\ref{fermion-chi coupling}) is of the general form
in the canonical eJBD formalism. 
We treat this in perturbation theory, regarding the constant $f$
in $F = 1 + f(\chi/M_{\rm P})^2$ as a small parameter.
To the leading $f-$order, the external source $\chi_0$ couples to fermions
in the following way 
\begin{eqnarray}
&&
{\cal L}_{\chi_0 \psi} 
\simeq
i f p_f\frac{ \chi_0 \partial_0 \chi_0 }{M_{\rm P}^2}\, \bar{\psi}\gamma^0 \psi
+ f p_Y \frac{\chi_0^2}{M_{\rm P}^2}\, m_{\psi} \bar{\psi} \psi
\,.
\label {fermion-chi coupling 2}
\end{eqnarray}

The second contribution in (\ref{fermion-chi coupling 2})
contains the trace of energy-momentum tensor of massive fermions.
The correction to the kinetic term given by the first contribution of
(\ref{fermion-chi coupling})
has an interesting structure:  it violates time-reversal symmetry
due to the time dependent background $\chi_0(t)$.
Note that field bilinear forms, $\chi_0 \partial_0 \chi_0$ and $\chi_0^2$,
couple to fermion pairs.
This implies that at one-loop level only diagrams of
the type shown in Fig(\ref{finite t diagram}) contribute
to the effective action at finite temperatures.
If one had included a linear term $\propto \chi/M_{\rm P}$ in $F$,
one would have derived three-point vertexes proportional to
$i p_f (\partial_0 \chi_0/M_{\rm P}) \bar{\psi}\gamma^0 \psi+
p_Y (\chi_0/M_{\rm P}) m_{\psi} \bar{\psi} \psi $,
however.
This would give additional contribution from diagrams
of vacuum polarization type in quantum electrodynamics.

\subsection
{\bf Effective action at finite temperatures}

We now calculate one-loop tadpole contribution.
The eJBD field $\chi$ is not in thermal equilibrium, but
fermions inside the loop diagram are assumed in thermal equilibrium
with the rest of environment particles including the same species
of fermions.
We use the Matsubara thermal Green's function \cite{landau-lifshitz},
which is derived from the Euclidean field theory limited to a finite Euclidean
time interval, $0 \sim \beta=1/T$.
The energy of thermal fermions is discretized in the energy unit
$\Delta \omega = 2\pi T$, and its relation
to the Minkowski energy $k_0$ is 
\begin{eqnarray}
&&
k_0 = i 2\pi T (n+ \frac{1}{2})
\,, \hspace{0.5cm}
n = 0, \pm 1, \pm 2, \cdots
\,.
\end{eqnarray}
Thus, the squared invariant is equal to 
\begin{eqnarray}
&&
k^2 = - (2\pi T)^2 (n+1/2)^2
- \vec{k}^2 \equiv - k_E^2 
\,.
\end{eqnarray}
The zero-temperature limit $\beta \rightarrow \infty $
is formally equivalent to the Euclidean field theory.
This means that perturbation formulas at finite temperatures
is nothing but the discretized sum (not a continuous integral) 
of formulas in the Wick-rotated perturbation theory

\vspace{0.5cm}
{\bf Kinetic-mixing correction}

\begin{figure*}[htbp]
 \begin{center}
 \epsfxsize=0.4\textwidth
 \centerline{\epsfbox{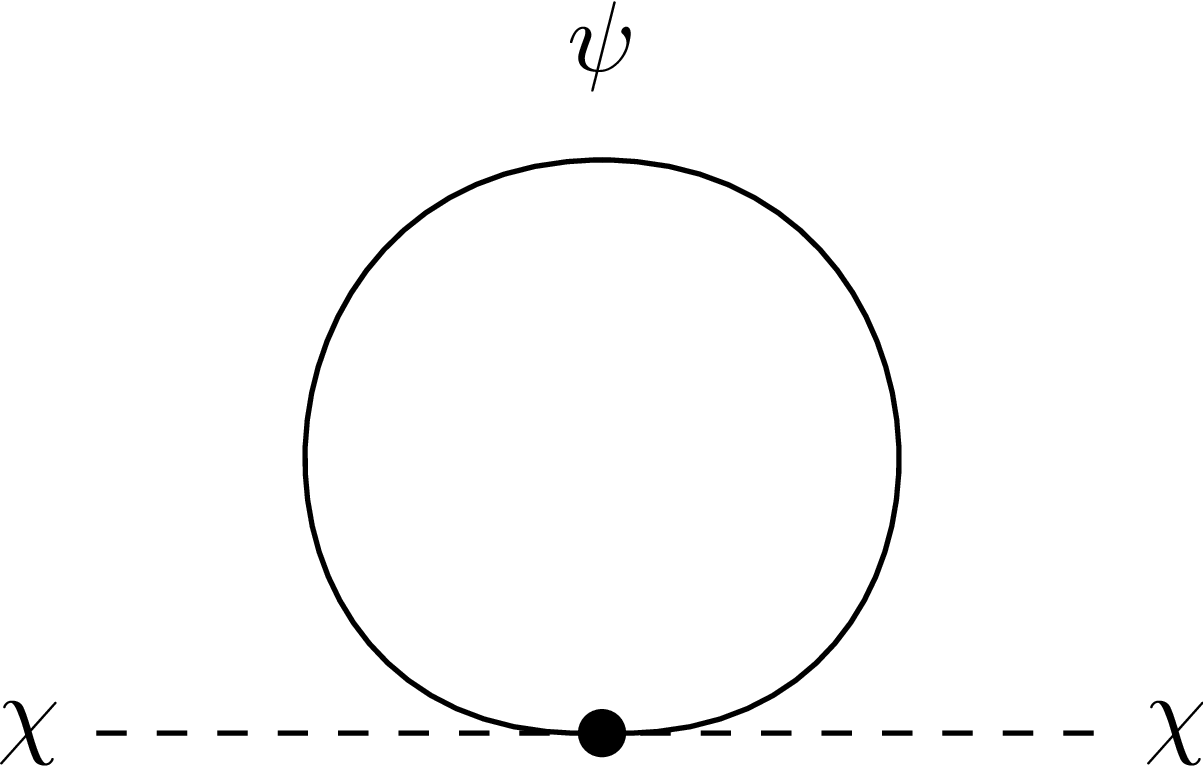}} \hspace*{\fill}\vspace*{3cm}
   \caption{
Feynman diagram of fermion loop.
}
   \label {finite t diagram}
 \end{center} 
\end{figure*}

The numerator of fermion propagator, with
the first vertex in (\ref{fermion-chi coupling 2}) of the diagram of Fig(\ref{finite t diagram}),
gives 
\begin{eqnarray}
&&
{\rm tr} \gamma^0 ( k \cdot \gamma + m_{\psi} ) = 4 k^0
= 4 i 2\pi T ( n + \frac{1}{2})
\,, \hspace{0.3cm}
n = 0, \pm 1, \pm 2, \cdots
\,,
\end{eqnarray}
contributing to a mixing term of the form,
$\chi_0 \partial_0 \chi$.
This is anti-symmetric in the exchange $n \rightarrow - n -1$,
while the denominator is symmetric.
Hence summation over $n$ gives vanishing contribution:
there is no kinetic correction at one-loop level.

\vspace{0.5cm}
{\bf Potential correction}

A straightforward calculation gives contribution of Fig(\ref{finite t diagram}) at finite
temperature, using the second vertex of (\ref{fermion-chi coupling 2}) 
in the diagram of Fig(\ref{finite t diagram}),
\begin{eqnarray}
&&
- i V^{(T)} = 
(-i)i (-1) p_Y f m_{\psi} \frac{\chi_0^2}{M_{\rm P}^2} i T\,
\int  \frac{d^3 k}{(2\pi)^{ 3}} \frac{ 4 m_{\psi}}{ - (2\pi T)^2 (n + \frac{1}{2})^2
- \vec{k}^2 - m_{\psi}^2}
\,.
\end{eqnarray}
We use the sum over discrete $n-$levels;
\begin{eqnarray}
&&
\sum_{n=1}^{\infty} \frac{1}{(2n-1)^2 + u^2} = \frac{\pi}{4u } \tanh \frac{\pi u}{2}
\,, \hspace{0.5cm}
u^2 = \frac{\vec{k}^2 + m^2}{ (2\pi T)^2}
\,.
\end{eqnarray}
The finite temperature effective potential is then
\begin{eqnarray}
&&
V^{(T)} = - 4 p_Y f \frac{\chi_0^2}{M_{\rm P}^2} \frac{m_{\psi}^2 T}
{(2\pi T)^2} \int   \frac{d^3 k}{(2\pi)^{ 3}}
\frac{\pi}{4u } \tanh \frac{\pi u}{2}
\,.
\end{eqnarray}

This involves quadratically divergent 3-momentum integral,
which is temperature dependent, the effective lagrangian being proportional to $ T^2 \chi_0^2$.
We adopt a practical rule that even temperature dependent mass terms
can be renormalized by subtraction.
The subtracted renormalized contribution then gives finite temperature
correction to the potential:
\begin{eqnarray}
&&
V^{(T)}_R = \frac{1}{4} p_Y f \frac{\chi_0^2}{M_{\rm P}^2} m_{\psi}^2 T^2\,
\int_0^{\infty} dv \frac{v^2}{\sqrt{v^2 + \mu_{\psi}^2}}
\left( 1 - \tanh (\frac{\pi}{2} \sqrt{v^2 + \mu_{\psi}^2})
\right)
\,, \hspace{0.3cm}
\mu_{\psi} = \frac{m_{\psi}}{2\pi T}
\,.
\end{eqnarray}
The integral is positive, and approaches $0.167$ in
the infinite temperature limit.

There are two important features in the renormalized effective
potential:
(1) correction to quadratic $\chi_0^2$ coefficient
of order, $m_{\psi}^2 T^2/M_{\rm P}^2$,
with $\chi_0/M_{\rm P} $ of order 5
at BBN epoch,
(2) its sign depends on the sign of $p_Y$ and
it can be either positive or negative.
$\chi$ mass corrected at finite temperatures can
be comparable to the Hubble rate in the temperature range
$T = O(m_{\psi})$.
This would modify the effective $\chi$ mass at BBN epoch
and  earlier. 
It is however expected that this mass contribution does not change
discussion on observational bounds at BBN.

During the most of radiation-dominated era including the electroweak epoch,
the squared mass $q^2 = O(V_0/M_{\rm P}^2) (t_0/t)$ is small.
Hence we took vanishing $q$ in the present calculation.
It exceeds, however, the electron-positron mass threshold; $q^2 > 4 m_e^2$
towards the inflationary epoch, and an important imaginary part emerges
above mass thresholds.

\vspace{0.5cm} \section {\bf Summary and outlook}

We have shown that introduction of standard particle physics
in cosmology based on extended Jordan-Brans-Dicke gravity is not a trivial matter.
The Weyl scaling from   the Jordan metric frame of
eJBD theory in which standard particle physics
lagrangian is naively introduced leads to
contradiction with observational and laboratory bounds on
time variation of standard model parameters.
Our new formulation is based on cosmology in the Einstein metric frame
in which interpretation of gravitational effects is evident.
It is necessary to extend basis of analysis prior to the Weyl scaling, and we
proposed to decompose standard particle physics
lagrangian into five independent gauge invariant pieces and to
assign different conformal factors to each of them.
This way we find a class of eJBD theories that evade all 
stringent observational bounds.

eJBD theories thus formulated are classified into two types,
depending on whether the effective potential for eJBD field
has a minimum at a finite field value or
the potential is slowly decreasing towards the field infinity.
The second type called type II we have considered in the present work 
is motivated by its natural relation to the cosmological constant problem,
and it  gives rise to sliding eJBD field at recent cosmic epochs, hence
time varying standard particle physics parameters.
Cosmological bounds in type \lromn2 models are sensitive to
the asymptotic behavior of the effective potential, but
insensitive to the  potential at smaller field values.
Five bounds on the power parameter $p_i$ derived in Section 4
have significant impact, but there are still a large parameter space of the allowd region.

We have found useful analytic solutions for time evolution
of eJBD scalar field and cosmic scale factor  supported by
numerical simulations.
The analysis based on this solution and 
detailed numerical simulations indicates that
the scalar field changes monotonically in radiation dominated epoch
including BBN time.
Late time behavior of three forms of energy densities is
culminated in illustrative numerical solution of Fig 2.

There exists new coupling of the form,
the eJBD field derivative $\partial \chi$ times 
 $ 3 j_B + j_L$ current.
This derivative coupling has interesting finite temperature effects, hitherto unnoticed.
This is a correction to kinetic terms of eJBD field.
The current form $ 3 j_B + j_L$
provides a new scenario to lepto-genesis, 
if the standard particle theory is extended to accommodate
heavy Majorana particles as in grand unified theories.
Coupling of eJBD field to fermion is found to give a new finite temperature
correction hitherto unnoticed, which may provide
interesting effects at higher temperatures than that of nucleo-synthesis.

Our pragmatic  approach has been phenomenological without deep fundamental
principle except that gauge invariance
and renormalizability of quantum field theory have been respected.
It would be interesting if a more fundamental approach such
as compactified superstring theory, \cite{string-motivated dilaton,
damour-polyakov}, leads to a theory
advocated by our approach.
As a more immediate application we mention
the possibility of eJBD theory giving a slow-roll inflation
and subsequent reheating.
It would be nice if one attributes the same eJBD scalar to
explanation of both inflation and dark energy problems.

\vspace{0.5cm} 
\appendix

\section{
Canonical Form of
Extended Jordan-Brans-Dicke Theory}

 It is the purpose of this Appendix to present canonically normalized
field theory lagrangian in eJBD gravity  that
incorporates the standard model of particle physics such that
the usual technical machinery of quantum field theory is made readily available.

In general, if there is a scalar that couples to the Ricci scalar non-minimally, a dynamical scalar degree of freedom emerges in the Einstein frame even if it has no kinetic term in the Jordan frame.
Such scalar field may couple to the gauge-singlet pieces of the Standard Model, which may cause time-dependence in various physical constants.

\subsection{General Framework}
We begin with the action defined in the Jordan frame, given by
\begin{align}
    S &=
    \int d^4x \sqrt{-g_J} \left[
        -\frac{M_{\rm P}^2}{2}F_{\rm g}(\phi)R_J
        + \frac{1}{2}F_{\rm{d}\phi}\partial^\mu\phi\partial_\mu\phi - V_{\rm J}(\phi)
    \right.\nonumber\\
    &\left.
        +F_{\rm dA}(\phi) {\cal L}_{\rm A,kin}
        +F_{\rm df}(\phi) {\cal L}_{\rm f,kin}
        +F_{\rm dH}(\phi) {\cal L}_{\rm H,kin}
        +F_{\rm Y}(\phi) {\cal L}_{\rm Y}
        -F_{\rm H}(\phi)V_{\rm J}(H)
    \right],
\end{align}
where
\begin{align}
    {\cal L}_{\rm A,kin} &= -\frac{1}{4}F^{\mu\nu} F_{J\mu\nu} - \sum_{a=1}^3\frac{1}{4} W^{a\mu\nu} W^a_{\mu\nu} - \sum_{A=1}^8\frac{1}{4}G^{A\mu\nu} G^A_{\mu\nu},\;\;\;
    {\cal L}_{\rm f,kin} = \sum_{\psi}\overline{\psi} i\gamma^\mu\nabla_\mu\psi,\\
    {\cal L}_{\rm H,kin} &=
    (D^\mu H)^\dagger (D_\mu H),\;\;\; 
    V_{\rm J}(H) =
    \lambda(|H|^2-v^2),\;\;\;
    {\cal L}_{\rm Y} =
    -y_u \overline{Q} H^c u - y_d \overline{Q} H d - y_e \overline{L} H e - y_\nu \overline{L}H^c \nu_R.
\end{align}
To stress that the metric, the Ricci scalar, and the scalar potentials are introduced
first in the Jordan frame, we denote in these formulas $g_{J\mu\nu}, R_J$, and $V_{\rm J}$, respectively.
Note that in the Jordan frame the cosmological constant may be included in
the potential energy $V_{\rm J}(\phi)$ of a quintessence field $\phi$.
The quintessence field or the dark energy field may consist of multiple components.

\subsection{Gravity sector}
Under the conformal transformation due to Weyl,
\begin{align}
    g_{J\mu\nu} &= F_{\rm g}^{-1} g_{\mu\nu},
    \label{eq:conformal transformation}
\end{align}
the Ricci scalar term is transformed to
\begin{align}
    F_{\rm g}^{-1} R_J &=
    R + 3g^{\mu\nu}\nabla_\mu\nabla_\nu \ln F_{\rm g} - \frac{3}{2}g^{\mu\nu}\partial_\mu\ln F_{\rm g}\partial_\nu\ln F_{\rm g}.
\end{align}
Thus, we obtain the relevant part of the action as 
\begin{align}
    S &=
    \int d^4x\sqrt{-g}\left[
        -\frac{M_{\rm P}^2}{2}R - \frac{3}{2}M_{\rm P}^2g^{\mu\nu}\nabla_\mu\nabla_\nu \ln F_{\rm g}
        + \frac{3}{4}M_{\rm P}^2g^{\mu\nu}\partial_\mu\ln F_{\rm g}\partial_\nu\ln F_{\rm g}+\frac{1}{2}\frac{F_{\rm{d}\phi}}{F_{\rm g}}g^{\mu\nu}\partial_\mu\phi\partial_\nu\phi-\frac{V_{\rm J}(\phi)}{F_{\rm g}^2}
    \right].\label{eq:S_g+L in E-frame}
\end{align}
We call the Einstein frame lagrangian when the gravity part of lagrangian density is
 given by $\sqrt{-g} R/16 \pi G_N$ as in this formula.

As in the text, we introduce the modulus field, $\chi$, of multiple component 
scalar fields.
The second term  of (\ref{eq:S_g+L in E-frame}) is a total derivative, hence
can be ignored.
The third and the forth terms of (\ref{eq:S_g+L in E-frame})
 may be recast to the form
of $\chi$ kinetic terms, $\partial^{\mu} \chi \partial_{\mu}\chi$:
\begin{eqnarray}
&&
\frac{3}{2}M_{\rm P}^2g^{\mu\nu}\partial_\mu\ln F_{\rm g}\partial_\nu\ln F_{\rm g}+\frac{F_{\rm{d}\phi}}{F_{\rm g}}g^{\mu\nu}\partial_\mu\phi\partial_\nu\phi = \partial^{\mu} \chi \partial_{\mu}\chi
\,.
\label {canonical ejbd field}
\end{eqnarray}
In order to introduce the canonical eJBD field $\chi$ this way, one has to impose some
condition.
We shall examine this condition for the simple case of a single-component field.
The left hand side of (\ref{canonical ejbd field}) may be written as\begin{eqnarray}
&&
\left( \frac{3}{2}M_{\rm P}^2 (\frac{\partial_{\phi} F_{\rm g} }{F_{\rm g} })^2
+ \frac{F_{\rm{d}\phi}}{F_{\rm g}} 
\right) g^{\mu\nu}\partial_\mu\phi\partial_\nu\phi
\,.
\end{eqnarray}
Assume that the bracketed part $\left( \cdots \right)$ of this equation
is a monotonic function of $\phi$. One can then define $\chi$ as
\begin{eqnarray}
&&
\partial \chi = \frac{2 }{ 3} \left( \frac{3}{2}M_{\rm P}^2 (\frac{\partial_{\phi} F_{\rm g} }{F_{\rm g} })^2
+ \frac{F_{\rm{d}\phi}}{F_{\rm g}} 
\right)^{3/2}\, \partial\phi 
\,.
\end{eqnarray}
If the bracketed function is not monotonic, correspondence between
$\phi$ and $\chi$ is not unique, and one has to define $\chi$ piecewise.

An effective  scalar potential of $\chi$ is defined by
\begin{align}
    V_{\rm eff} (\chi) &= \frac{V_{\rm J}(\phi)}{F^2_g},
\end{align}
and $\chi$ plays the role of dark energy at later times.
Regardless of the specific form of $V_{\rm J}(\chi)$\footnote{This means $V_{\rm J}(\phi(\chi))$.}, the effective theory of gravity and $\chi$ is thus given by
\begin{align}
    S &=
    \int d^4x\sqrt{-g}\left[
        -\frac{M_{\rm P}^2}{2}R + \frac{1}{2}g^{\mu\nu}\partial_\mu\chi\partial_\nu\chi 
- V_{\rm eff}(\chi)
    \right].
\end{align}
The gravity and the scalar part in the Jordan frame is thus equivalent to
the Einstein gravity added by the canonical form of scalar field lagrangian
after Weyl scaling.
We take $V_{\rm eff}(\chi)$  to be a quartic function of $\chi$, while
$F_{\rm g}$ to be a quadratic function.

\subsection{Gauge sector}
For simplicity, we consider only the QED part in the gauge kinetic sector.
In the Einstein frame, the action becomes
\begin{align}
    S &=
    \int d^4 x\sqrt{-g}\frac{F_{\rm dA}}{F_{\rm g}^2} {\cal L}_{\rm A,kin}
    =
    \int d^4 x\sqrt{-g}\left[
        -\frac{1}{4}F_{\rm dA}g^{\mu\alpha}g^{\nu\beta} F_{\mu\nu} F_{\alpha\beta}
    \right].
\end{align}
Note that $\partial_\mu (=\partial/\partial x^\mu)$ does not change under the conformal transformation, but $\partial^\mu$ changes as $\partial^\mu = g^{\mu\nu}\partial/\partial x^\nu$.
The gauge field may be canonically normalized by rescaling as $A_\mu \to F_{\rm dA}^{-1/2} A_\mu$, and thus
\begin{align}
    S = \int d^4x\sqrt{-g}\left[
        -\frac{1}{4}g^{\mu\alpha}g^{\nu\beta}F_{\mu\nu}F_{\alpha\beta}
    \right].
\end{align}

\subsection{Fermion sector}
Fermion kinetic term involves the spin connection whose convenient expression is given by
\begin{align}
    \omega_\mu{}^{ab} &= -e^{\nu b}\nabla_\mu e^a_\nu = -e^{\nu b}(\partial_\mu e^a_\nu - \Gamma^\rho_{\mu\nu} e^a_\rho),
\end{align}
where the covariant derivative acts on $e^a_\mu$ by supposing that $e^a_\mu$ is a covariant vector, which is indeed the case as $e^a_\mu$ is a set of covariant vectors, namely, $e^0_\mu, e^1_\mu, ...$
It turns out that this is equivalent to the metric postulate, i.e., $\nabla_\alpha g_{\mu\nu}=0$.
For the FLRW metric, the frame field components are given by
\begin{align}
    &
    e^a_\mu = {\rm diag}(1,a,a,a),\;\;\;
    e^\mu_a = {\rm diag}(1,a^{-1},a^{-1},a^{-1}),\\
    &
    e_{\mu a} = \eta_{ab}e^b_\mu = {\rm diag}(1,-a,-a,-a),\;\;\;
    e^{\mu a} = \eta^{ab}e^\mu_b = {\rm diag}(1,-a^{-1},-a^{-1},-a^{-1}).
\end{align}
The affine connection $\Gamma^\rho_{\mu\nu}$ is given by 
\begin{align}
    &
    \Gamma^0_{00} = \Gamma^0_{0i} = \Gamma^0_{i0} = \Gamma^i_{00} = \Gamma^i_{jk} = 0,\;\;\;
    \Gamma^0_{ij} = a^2H\delta_{ij},\;\;\;
    \Gamma^i_{0j} = \Gamma^i_{j0} = H \delta^i_j,
\end{align}
where $H=\dot{a}/a$.
From these results, we end up with
\begin{align}
    \omega_0{}^{ab} = 0, \;\;\; \omega_i{}^{ab} = aH(\delta^a_i\delta^b_0 - \delta^a_0\delta^b_i)
\end{align}
for $i=1,2,3$, which yields
\begin{align}
    \gamma^\mu\nabla_\mu &=
    \gamma^\mu\left(
        \partial_\mu + \frac{1}{4}\omega_{\mu ab}\gamma^{ab}
    \right)
    =
    \gamma^\mu\partial_\mu + \frac{3}{2}H\gamma^0.
\end{align}

The conformal transformation given by Eq. (\ref{eq:conformal transformation}) can be rewritten in terms of the frame field as
\begin{align}
    e^a_{J\mu} &= F_{\rm g}^{-1/2}e^a_\mu.
\end{align}
Under the conformal transformation the spin connection becomes
\begin{align}
    \omega_\mu{}^{ab}(e_J) &=
    \omega_\mu{}^{ab}(e) - \frac{1}{2}(e^a_\mu e^b_\nu - e^b_\mu e^a_\nu)\partial^\nu\ln F_{\rm g}.
\label {origin of derivative coupling}
\end{align}
Noticing $\gamma^\mu(e_J) = e^\mu_{Ja} \gamma^a = F_{\rm g}^{1/2}\gamma^\mu(e)$, we obtain the action as
\begin{align}
    S &=
    \int d^4 x\sqrt{-g} F_{\rm g}^{-2}F_{\rm df}{\cal L}_{\rm F,kin}
    =
    \int d^4 x\sqrt{-g}F_{\rm g}^{-3/2}F_{\rm df}\times \overline{\psi}i\gamma^\mu\left(
        \partial_\mu - ig_A q_A A_\mu - \frac{3}{4}\partial_\mu \ln F_{\rm g}
    \right)\psi,
\end{align}
where $\gamma^\mu = \gamma^\mu(e)$ and the gauge field is explicitly included.
The sum over all fermion species for $\psi$ is implicitly assumed.
We may rescale the fermions as $\psi \to F_{\rm g}^{3/4}F_{\rm df}^{1/2}\psi$ to eliminate the term having the piece of $-(3/4)\partial_\mu\ln F_{\rm g}$.
Instead, a term proportional to $\partial_\mu \ln F_{\rm df}$ appears.
The canonical form of the fermion kinetic term thus becomes
\begin{align}
    S &=
    \int d^4x \sqrt{-g}\overline\psi i\gamma^\mu \left(
        \partial_\mu - ig_A q_A F_{\rm dA}^{-1/2}A_\mu + \frac{1}{2}\partial_\mu \ln F_{\rm df}
    \right)\psi,
\end{align}
where we have also rescaled the gauge field as $A_\mu \to F_{\rm dA}^{-1/2}A_\mu$ as performed in the previous section to make the gauge kinetic term canonical.
It turns out that the effective gauge coupling in the Einstein frame is given by
\begin{align}
    g_{A,{\rm eff}} &= F_{\rm dA}^{-1/2} g_A.
\end{align}
We also find that fermions receive a universal coupling given by
\begin{align}
    \frac{i}{2}(\partial_\mu\ln F_{\rm df})\times \overline\psi \gamma^\mu \psi
= \frac{i}{2}(\partial_{\chi} \ln F_{\rm df})\, 
\partial_\mu \chi  \overline{\psi} \gamma^\mu \psi.
\label{eq:Sdf}
\end{align}

\subsection{Higgs sector}

Under the conformal transformation, the Higgs kinetic term becomes
\begin{align}
    S &=
    \int d^4x\sqrt{-g}\frac{F_{\rm dH}}{F_{\rm g}^2}F_{\rm g} g^{\mu\nu}(D_\mu H)^\dagger (D_\nu H).
\end{align}
Writing $f_{\rm dH}^2\equiv F_{\rm dH}/F_{\rm g}$ and rescaling $H \to f_{\rm dH}^{-1}H$, the action can be rewritten as
\begin{align}
    S &=
    \int d^4 x\sqrt{-g}\left[
        g^{\mu\nu}(D_\mu H)^\dagger (D_\nu H) 
        +g^{\mu\nu}\left\{
            (\partial_\mu\ln f_{\rm dH})(\partial_\nu\ln f_{\rm dH})+2\nabla_\mu\nabla_\nu\ln f_{\rm dH}
        \right\}|H|^2
    \right],
    \label{eq:SdH}
\end{align}
where we have swapped $D_\mu$ with $\nabla_\mu$ whenever possible and have integrated by parts with $\nabla_\alpha g_{\mu\nu}=0$ to obtain the last term.
Notice that the last term in Eq. (\ref{eq:SdH}) is an effective mass term for the Higgs boson.
So, the effective Higgs potential in the Einstein frame, $V_{\rm eff}(H)$, may be written as 
\begin{align}
    V_{\rm eff}(H) &= F_{\rm H} F_{\rm g}^{-2}V_{\rm J}(H) - g^{\mu\nu}\left\{
        (\partial_\mu\ln f_{\rm dH})(\partial_\nu\ln f_{\rm dH})+2\nabla_\mu\nabla_\nu\ln f_{\rm dH}
    \right\}|H|^2.
\label {higgs canonical potential}
\end{align}

Existence of kinetic mixing $\partial^{\mu} \chi \partial_{\mu} \chi |H|^2$
times a function of $\chi$ field in the second term of this formula
 is an inevitable consequence of
the canonical transformation we performed.
This term may be further recast to the form,
\begin{eqnarray}
&&
- \left(
(\partial_{\chi} \ln f_{\rm dH} )^2 + 2 \partial^2_{\chi} \ln f_{\rm dH} \right)
\partial^{\mu} \chi
\partial_{\mu} \chi + 2 \partial_{\chi} \ln f_{\rm dH}\, \partial^{\mu}\partial_{\mu} \chi
\,.
\end{eqnarray}
This contribution along with a new fermion coupling (\ref{eq:Sdf})
may greatly change finite-temperature corrections to the effective
$\chi$ potential compared without these.

\subsection{Yukawa coupling part}

Noticing that the canonically normalized Higgs ($H$) and fermions ($\psi$) are defined by rescaling the original field as 
\begin{align}
    H \to (F_{\rm g}/F_{\rm dH})^{1/2} H,
    \;\;\;
    \psi \to F_{\rm g}^{3/4}F_{\rm df}^{1/2}\psi,
\end{align}
we obtain the Einstein-frame Yukawa couplings as
\begin{align}
    S &= \int d^4x \sqrt{-g}F_{\rm g}^{-2}F_{\rm Y} {\cal L}_{\rm Y}\nonumber\\
    &=
    \int d^4x\sqrt{-g} (F_{\rm g}^{-2}F_{\rm Y}) (F_{\rm g}^{3/4}F_{\rm df}^{1/2})^2 (F_{\rm g}/F_{\rm dH})^{1/2}
    \left(
        -y_u \overline Q H^c u - y_d \overline Q H d - y_e \overline L H e - y_{\nu} \overline L H^c 
\nu_R
    \right).
\end{align}
We wrote Yukawa coupling for the first generation of quarks and leptons,
and contributions from other two generations should be added similarly.
Neutrinos are known to be massive, and for explanation of the smallness of
neutrino masses it is often assumed that the seesaw mechanism
between right-handed heavy Majorana leptons $\nu_R$
(to be presumably denoted by $N_R$ for a better clarity to differentiate from $\nu_L$
of two Majorana particles with huge mass differences) and ordinary left-handed neutrinos
are present.
We do not commit to the interesting case of Majorana neutrino $\nu_L$ alone,
and leave the other possibility of Dirac neutrino masses.
For both cases the effective Yukawa coupling is given by
\begin{align}
    y_{f,{\rm eff}} &= F_{\rm Y}F_{\rm df} F_{\rm dH}^{-1/2} y_f \,.
\end{align}

\section{Techniques for Numerical Computation}

The equation of motion of eJBD field $\chi$ is not solvable by analytic means and
we need to numerically analyze the dynamics of time evolution.
We explain here some useful techniques for numerical analysis
and present time evolution after BBN, those of
eJBD field, radiation and matter energy density.
For notational simplicity we write $\chi$ potential as $V$
replacing $V_{\rm eff}$ in this Appendix.

In many cases, it is easier to solve the equation of motion 
in terms of a rescaled time variable. 
In the present case, we may use the number of e-folds, $N$, introduced as
$N = \ln(a/a_0)$, where $a_0$ is the scale factor of the present Universe and is taken as $a_0=1$ in the following.
Accordingly, the derivatives with respect to $t$ become $\frac{d}{dt} = H\frac{d}{dN},
\frac{d^2}{dt^2} = \dot{H}\frac{d}{dN} + H^2\frac{d^2}{dN^2}$, with $H = H(t)$ the Hubble rate.

The Hubble rate is given by
\begin{align}
    &
    3M_{\rm P}^2 H^2 
    = \rho_r + \rho_m + \rho_\chi= \rho_{r,0} a^{-4} + \rho_{m,0} a^{-3} + \frac{1}{2}\dot{\chi}^2 + V(\chi),
\end{align}
where, from the Planck observation \cite{Planck:2018vyg}, the energy densities of radiation and matter measured at $a=a_0$ are given by
\begin{align}
    &
    \rho_{r,0} = 3.366\times 10^{-51}~{\rm GeV}^4,\;\;\;
    \rho_{m,0} = 1.147\times 10^{-47}~{\rm GeV}^4.
\end{align}
Writing $d\chi/dN\equiv v$ and thus $\dot\chi=Hv$, we may rewrite $H$ as
\begin{align}
   3M_{\rm P}^2 H^2 &= \frac{\rho_{r,0}e^{-4N}+\rho_{m,0}e^{-3N}+V}{1-v^2/6M_{\rm P}^2}. \label{eq:Hubble}
\end{align}
From Eq. (\ref{eq:Hubble}), we find
\begin{align}
    \rho_\chi &= \frac{1}{1-v^2/6M_{\rm P}^2}\left(
        \frac{\rho_r+\rho_m}{6M_{\rm P}^2}v^2+V
        \right).
\end{align}

In solving the equation of motion, we also need $\dot H$, which is given by
\begin{align}
    6M_{\rm P}^2 \dot{H}H &= \dot\rho_r + \dot\rho_m + \dot\rho_\chi = -H(4\rho_r+3\rho_m) -3H^3 v^2,
\end{align}
where we have used
\begin{align}
    &
    \dot\rho_r = -4H\rho_r,\;\;\;
    \dot\rho_m = -3H\rho_m.
\end{align}
Also, by noticing that $\ddot\chi + \partial_\chi V = -3H\dot\chi$, we have
\begin{align}
    \dot\rho_\chi = \dot\chi(\ddot\chi+\partial_\chi V) = -3H^3v^2.
\end{align}

Therefore, the ordinary differential equations we solve are
\begin{align}
    &
    \frac{d\chi}{dN} = v,\;\;\;
    \frac{dv}{dN} = -\frac{\dot H+3H^2}{H^2}v - \frac{\partial_\chi V}{H^2}
    \label{eq:ODE1}
\end{align}
together with
\begin{align}
    &
    H^2 = \frac{1}{3M_{\rm P}^2}\frac{\rho_{r,0}e^{-4N}+\rho_{m,0}e^{-3N}+V}{1-v^2/6M_{\rm P}^2},\;\;\;
    \dot H = -\frac{2}{3}\frac{\rho_{r,0}}{M_{\rm P}^2}e^{-4N} - \frac{1}{2}\frac{\rho_{m,0}}{M_{\rm P}^2}e^{-3N} - \frac{1}{2} \frac{H^2v^2}{M_{\rm P}^2}.
    \label{eq:ODE2}
\end{align}

\begin{figure*}[htbp]
\includegraphics[width=.8\textwidth,angle=90]{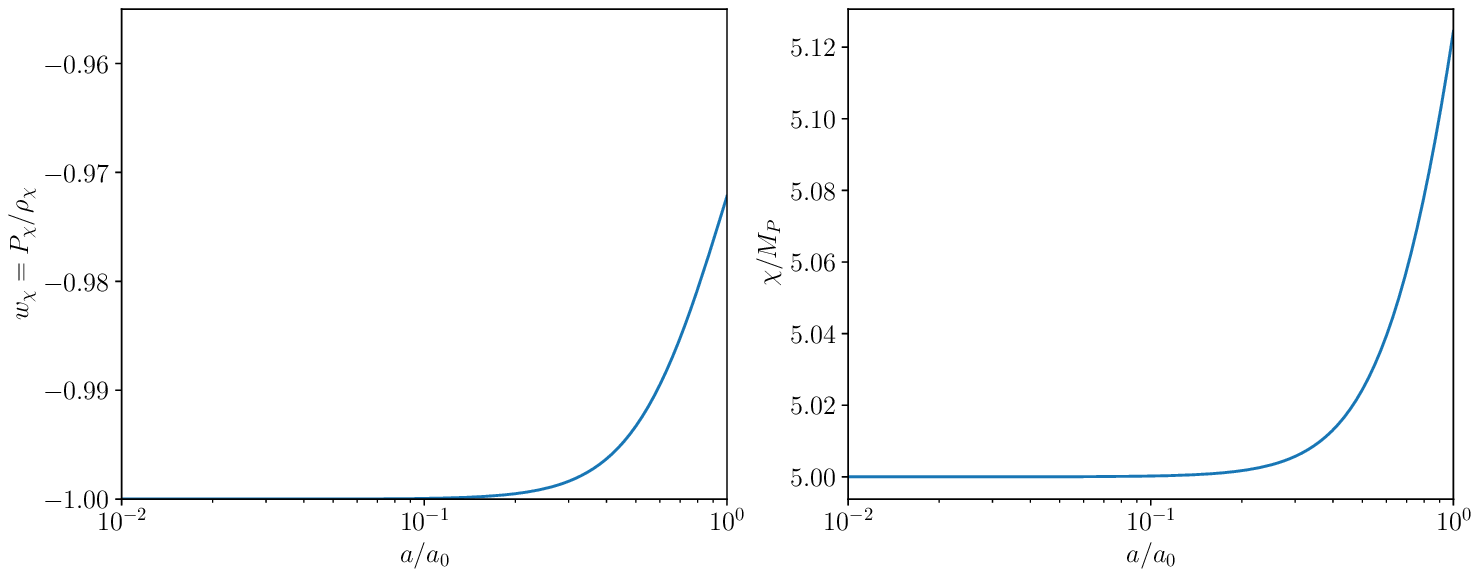}
 \begin{center}
   \caption{
Time evolution of equation of state parameter $w = p/\rho$
and eJBD field $\chi$
at recent epochs of $z= 10^{2} \sim 0$.
Assumed parameters are the same as in Fig(\ref{time-evolution_fig1}).}

   \label {time-evolution_fig2}
 \end{center} 
\end{figure*}

Solving Eqs. (\ref{eq:ODE1}) and (\ref{eq:ODE2}), we have evaluated the energy fractions, the equation-of-state parameter, and the field excursion of $\chi$ 
in Fig(\ref{time-evolution_fig1}) and 
(\ref{time-evolution_fig2}), respectively, 
where the initial condition has been set as $\chi/M_{\rm P}=5, v=0$ at $a=10^{-10}$ corresponding to roughly the onset of BBN.
The scalar potential is taken to be
\begin{align}
    V(\chi) &= \frac{V_{\rm J} (\chi)}{F_{\rm g}^2(\chi)}
    = \frac{V_0(d_4(\chi/M_{\rm P})^4 + d_2 (\chi/M_{\rm P})^2 + 1)}{(1+f(\chi/M_{\rm P})^2)^\eta}.
\end{align}
In Fig(\ref{time-evolution_fig1}), we have defined $\Omega_i \equiv \rho_i/\rho_{\rm cr}$ for $i=r,m,\chi$ where $\rho_{\rm cr}\equiv 3M_{\rm P}^2H^2$ at a given time.
At $a=a_0$, we find
\begin{align}
    \Omega_{\chi,0} &= 0.681,
\end{align}
which is consistent with the Planck data \cite{Planck:2018vyg}.
At the BBN, the energy fraction of $\chi$ is obtained as
\begin{align}
    \Omega_{\chi,{\rm BBN}} &= 7.57\times10^{-37},
\end{align}
which is sufficiently small.
Such additional component in the energy density is constrained in terms of $\Delta N_{\rm eff}$, the deviation from the number of the effective neutrino degrees $N_{\rm eff} = 3.046$, which at 68\% CL, we have $N_{\rm eff} = 3.27\pm 0.15$ \cite{Planck:2018vyg}.
The allowed amount of $\Omega_{\chi,{\rm BBN}}$ is estimated by \cite{Ferreira:1997au}
\begin{align}
    \Omega_{\chi,{\rm BBN}} \lesssim \frac{3}{4}\frac{7\Delta N_{\rm eff}/4}{10.75+\Delta N_{\rm eff}/4}\sim 0.043,
\end{align}
where we have used $\Delta N_{\rm eff} = 0.374$ \cite{Planck:2018vyg}.

For the equation of state parameter defined by $w_\chi\equiv P_\chi/\rho_\chi$ where the pressure $P_\chi$ is given by $P_\chi \equiv \dot\chi^2/2 - V$, we find
\begin{align}
    w_\chi &= -0.972
\end{align}
at $a=a_0$, which satisfies the Planck data $w_\chi < -0.95$ given in Eq. (51) of Ref. 
\cite{Planck:2018vyg}.

\subsection*{Acknowledgement}
We appreciate P. Fayet for pointing out an error in the original version of
this work.
The work is in part supported by JSPS KAKENHI Grant Nos. ~19H01899 (KK and KO),
 21H01107 (KO), and 21K03575 (MY).


\end{document}